%% file: CI-LMN_arXiv_v3_FrewerKhujadze.tex
\documentclass[11pt,a4,twoside]{article}
\usepackage[dvips,a4paper,left=2.5cm,right=2.5cm,top=2.5cm,bottom=2.5cm,headsep=1em]{geometry}
\usepackage{fancyhdr}
\usepackage{titlesec}
\usepackage[latin1]{inputenc}
\usepackage{amsmath,amsfonts,amssymb,array,amsthm}
\usepackage{rotating}
\usepackage[font={small}]{caption}
\usepackage{natbib}
\usepackage{lmodern,slantsc}
\usepackage{multirow}
\usepackage{chngcntr}
\usepackage{placeins}
\usepackage{caption}
\usepackage{enumitem}
\usepackage{maplestd2e}

\DefineParaStyle{Maple Output} \DefineParaStyle{Maple Output}
\DefineCharStyle{2D Math} \DefineCharStyle{2D Output}
\usepackage[breaklinks,
colorlinks=true, linkcolor=blue, citecolor=black, urlcolor=blue,
pdfborder={0 0 0}, pdfpagelabels]{hyperref}
\usepackage{tikz}
\newcommand{\ExternalLink}{%
\tikz[x=1.2ex, y=1.2ex, baseline=-0.05ex]{%
\begin{scope}[x=1ex, y=1ex]
\clip (-0.1,-0.1) --++ (-0, 1.2) --++ (0.6, 0) --++ (0, -0.6) --++
(0.6, 0) --++ (0, -1); \path[draw, line width = 0.5, rounded
corners=0.5] (0,0) rectangle (1,1);
\end{scope}
\path[draw, line width = 0.5] (0.5, 0.5) -- (1, 1); \path[draw,
line width = 0.5] (0.6, 1) -- (1, 1) -- (1, 0.6);}}
\def\vx{\mathbf x}
\def\vy{\mathbf y}

\def\vJ{\mathbf J}

\def\v0{\boldsymbol{0}}

\newlength{\FigureHeight}
\newlength{\FigureHeightHalf}

\numberwithin{equation}{section}
\titleformat{\section}
{\large\bfseries}{\thetitle.}{0.5em}{}
\titleformat{\subsection}
{\normalfont\bfseries}{\thetitle.}{0.5em}{}
\titleformat{\subsubsection}
{\normalfont\itshape}{\normalfont\thetitle.}{0.5em}{}
\begin{document}

\newgeometry{left=2.5cm,right=2.5cm,top=2.5cm,bottom=2.0cm,headsep=1em}

\title{\vspace{-2.0em} Conformal invariance and the Lundgren-Monin-Novikov equations for vorticity fields in 2D turbulence: Refuting a recent claim}
\author{Michael Frewer$\,^1$\thanks{Email address for correspondence:
frewer.science@gmail.com}$\:\,$ \& George Khujadze$\,^2$ \\ \\
\small $^1$ Heidelberg, Germany\\
\small $^2$ Chair of Fluid Mechanics, Universit\"at Siegen, 57068
Siegen, Germany}
\date{{\small\today}}
\clearpage \maketitle \thispagestyle{empty}

\vspace{-2.0em}\begin{abstract}

\noindent The recent claim by Grebenev~{\it et~al.}~[\href{https://doi.org/10.1088/1751-8121/aa8c69}{J.~Phys.~A:~Math.~Theor.$\,$50,~435502~(2017)}] that\linebreak[4] the inviscid 2D Lundgren-Monin-Novikov (LMN) equations on a zero vorticity characteristic naturally would reveal local conformal invariance when only analyzing these by means of a classical Lie-group symmetry approach, is invalid and will be refuted in the present comment. To note is that within this comment the (possible) existence of conformal invariance in 2D turbulence is not questioned, only the conclusion as is given in \cite{Grebenev17} and their approach how this invariance was derived is what is being criticized and refuted herein. In~fact, the algebraic derivation for conformal invariance of the 2D LMN vorticity equations in\linebreak[4] \cite{Grebenev17} is flawed. A key constraint of the LMN equations has been wrongly transformed. Providing the correct transformation instead will lead to a breaking of the proclaimed conformal group. The corrected version of \cite{Grebenev17} just leads to a globally constant scaling in the fields and not to a local one as claimed. In consequence,\linebreak since in \cite{Grebenev17} only the first equation within the infinite and unclosed\linebreak[4] LMN chain is considered, also different Lie-group infinitesimals for the one- and two-point probability density functions (PDFs) will result from this correction, replacing thus the misleading ones proposed.

\vspace{0.5em}\noindent{\footnotesize{\bf Keywords:} {\it Statistical Physics, Conformal Invariance, Turbulence, Probability Density Functions, Lie Groups, Symmetry Analysis, Integro-Differential Equations,
Closure Problem}}
\end{abstract}

\section{Summary of the key results obtained in \cite{Grebenev17}\label{Sec1}}

Considered is the first equation in the unclosed chain of the inviscid 2D Lundgren-Monin-Novikov (LMN) vorticity equations (Eq.$\,$[3])
\begin{multline}
\frac{\partial f_1(\vx,\omega,t)}{\partial t}-\frac{\partial}{\partial x^1}\int d^2\vx^\prime d\omega^\prime \omega^\prime\frac{x^2-x^{\prime 2}}{2\pi|\vx-\vx^\prime|^2}f_2(\vx,\omega,\vx^\prime,\omega^\prime,t)\\
+\frac{\partial}{\partial x^2}\int d^2\vx^\prime d\omega^\prime \omega^\prime\frac{x^1-x^{\prime 1}}{2\pi|\vx-\vx^\prime|^2}f_2(\vx,\omega,\vx^\prime,\omega^\prime,t)=0,\hspace{0.60cm}
\label{180130:2115}
\end{multline}
describing the dynamics of the 1-point probability density function (PDF) $f_1$ in terms of the (unclosed) 2-point PDF $f_2$, where $\omega$ and $\omega^\prime$ denote the sample space variables of the single vorticity component at the space-time points $(\vx,t)$ and $(\vx^\prime,t)$, respectively. This equation \eqref{180130:2115} is supplemented by a normalization constraint for each of the two PDFs (Eq.$\,$[4])
\begin{equation}
\int d\omega f_1=1,\qquad \int d\omega^\prime f_2=f_1.\label{180130:2158}
\end{equation}
Relations \eqref{180130:2115} and \eqref{180130:2158} form the complete set of equations which then were subjected to a systematic Lie-group symmetry analysis in \cite{Grebenev17}. By introducing the following vector of independent variables (Eq.$\,$[5])
\begin{equation}
(y^0,\vy)=(t,\vy)=(t,\vx,\omega,\vx^\prime,\omega^\prime)=(t,x^1,x^2,\omega,x^{\prime 1},x^{\prime 2},\omega^\prime),\label{180201:1054}
\end{equation}
and by using this $\vy$ notation interchangeably with the original $\vx$ notation, this governing system\pagebreak[4]

\restoregeometry

\noindent of equations \eqref{180130:2115}-\eqref{180130:2158} can be equivalently rewritten as (Eqs.$\,$[6-7])
\begin{align}
\mathsf{E}_1\!:&\quad \frac{\partial J^0}{\partial y^0}+ \frac{\partial J^1}{\partial y^1}+\frac{\partial J^2}{\partial y^2}=0,\label{180131:0807}\\[0.5em]
\mathsf{E}_2\!:&\quad J^1+\frac{1}{2\pi}\int d^2\vx^\prime d\omega^\prime \omega^\prime\frac{x^2-x^{\prime 2}}{|\vx-\vx^\prime|^2}f_2=0,\label{180201:1004}\\[0.5em]
\mathsf{E}_3\!:&\quad J^2-\frac{1}{2\pi}\int d^2\vx^\prime d\omega^\prime \omega^\prime\frac{x^1-x^{\prime 1}}{|\vx-\vx^\prime|^2}f_2=0,\label{180201:1005}\\[0.5em]
\mathsf{E}_4\!:&\quad 1-\int d\omega f_1=0,\label{180201:1006}\\[0.5em]
\mathsf{E}_5\!:&\quad f_1-\int d\omega^\prime f_2=0,\label{180201:1007}
\end{align}
where $J^0:=f_1$. The Lie-group symmetry analysis for the above system was performed successively in \cite{Grebenev17}, first for equation $\mathsf{E}_1$, then by including $\mathsf{E}_2$ and $\mathsf{E}_3$ into the analysis, to then finally restrict the obtained symmetry result by $\mathsf{E}_4$ and $\mathsf{E}_5$. To note is that only the first step, i.e.~the symmetry analysis for $\mathsf{E}_1$, was discussed generally, while all subsequent steps were performed under a specific Lie-point symmetry ansatz to explicitly bring forward a local conformal invariance for this system.

In infinitesimal form, the most general Lie-point symmetry admitted by $\mathsf{E}_1$ \eqref{180131:0807}, being itself an equation in continuity form with four independent and three dependent variables, is given
as~(Eqs.$\,$[27-30])
\begin{align}
X=&\;\xi^0(t,\vx,\omega)\frac{\partial}{\partial t}+\xi^1 (t,\vx,\omega)\frac{\partial}{\partial x^1}+\xi^2 (t,\vx,\omega)\frac{\partial}{\partial x^2}+\xi^3 (\omega)\frac{\partial}{\partial \omega}\nonumber\\[0.5em]
&\; +\eta^0(t,\vx,\omega,\vJ)\frac{\partial}{\partial J^0}+\eta^1(t,\vx,\omega,\vJ)\frac{\partial}{\partial J^1}+\eta^2(t,\vx,\omega,\vJ)\frac{\partial}{\partial J^2},\label{180201:0924}
\end{align}
with
\begin{equation}
\eta^i(t,\vx,\omega,\vJ)=a^i_k(t,\vx,\omega)J^k+b^i(t,\vx,\omega),\quad\; i,k=0,1,2,\label{180201:1224}
\end{equation}
where the coefficients $a^i_k$ have the specified form
\begin{equation}
a^i_k=\xi^i_k-\delta^i_k\Big(\xi^0_0+\xi^1_1+\xi^2_2+C(\omega)\Big),\quad\;\; \xi^i_k:=\frac{\partial \xi^i}{\partial y^k},\label{180201:0930}
\end{equation}
and where the $b^i$ are arbitrary solutions of $\mathsf{E}_1$ \eqref{180131:0807}. Note that while the three infinitesimals $\xi^0$, $\xi^1$ and $\xi^2$ are arbitrary (1-point) space-time functions, the generating infinitesimal for the vorticity $\xi^3$, however, is independent of space and time; it is an arbitrary function only of its own defining variable $y^3=\omega$. This result stems from the fact that the variable $y^3=\omega$ is not explicit in equation $\mathsf{E}_1$ \eqref{180131:0807} with the effect then that a symmetry analysis identifies it as a hidden parameter that only can be arbitrarily re-parametrized. In \eqref{180201:0930}, function~$C$ is also only a function of $\omega$ not depending on space and time. The above result \eqref{180201:0924}-\eqref{180201:0930} has been independently validated by using the computer algebra package DESOLV-II of \cite{Vu12},
matching the result given in \cite{Grebenev17} by Eqs.$\,$[27-30], up to the minor misprint\footnote{If claimed not to be a misprint, then it is definitely a mistake in \cite{Grebenev17} to denote the dependencies of the infinitesimals in Eqs.$\,$[27-30] with $\vy$ instead of $(\vx,\omega)$. The reason is that if the $J^i$ are formally identified as functions of $\vy$, then equation $\mathsf{E}_1$ \eqref{180131:0807} has to be augmented by the 9 constraints $J^i_k=0$, for $k=4,5,6$, to indicate and to provide the relevant information that all $J^i$ are 1-point and not 2-point functions.\label{fn1}} in the dependencies of the infinitesimals (instead of $\vy$ only $\vx,\omega$), the missing constraint $\xi^3_0=0$ in Eq.$\,$[30], and that $C\equiv C(\omega)$ --- see Appendix \ref{Sec.A} for an explicit proof of this result \eqref{180201:0924}-\eqref{180201:0930}.

\noindent
Based on this result \eqref{180201:0924}-\eqref{180201:0930} for $\mathsf{E}_1$, the second step in \cite{Grebenev17} includes the equations $\mathsf{E}_2$ \eqref{180201:1004} and $\mathsf{E}_3$ \eqref{180201:1005} into the symmetry analysis, however, not generally, but rather with~the following specifically chosen ansatz for the infinitesimals (Eqs.$\,$[A.35-A.38])\footnote{Note that the infinitesimal for the time variable $\xi^0$ in \cite{Grebenev17} has been ultimately put to zero,\linebreak[4] not during the symmetry analysis itself, which therein was explicitly performed in Appx.$\,$A, but later when discussing the result in Sec.$\,$3 on p.$\,$8; see Eq.$\,$[33]. Thus, for convenience, $\xi^0$ is considered herein throughout~as~zero.}
\begin{align}
\xi^0 &= 0,\label{180201:1057}\\[0.5em]
\xi^1 &= c^{11}(\vx)x^1 +c^{12}(\vx)x^2+d^1(\vx),\label{180201:1058}\\[0.5em]
\xi^2 &= c^{21}(\vx)x^1 +c^{22}(\vx)x^2+d^2(\vx),\label{180201:1059}\\[0.5em]
\xi^4 &= c^{11}(\vx)x^{\prime 1} +c^{12}(\vx)x^{\prime 2}+d^1(\vx),\label{180201:1100}\\[0.5em]
\xi^5 &= c^{21}(\vx)x^{\prime 1} +c^{22}(\vx)x^{\prime 2}+d^2(\vx),\label{180201:1101}
\end{align}
along with the constraints (Eq.$\,$[A.40] leading to [A.41])
\begin{equation}
c^{22}(\vx)=c^{11}(\vx)\quad\text{and}\quad c^{21}(\vx)=-c^{12}(\vx),\label{180201:1125}
\end{equation}
which, as an overall result, already represents the structure of a local conformal invariance for the combined system $\mathsf{E}_1$-$\mathsf{E}_3$. Note that according to notation \eqref{180201:1054}, the functions $\xi^4$ \eqref{180201:1100} and $\xi^5$~\eqref{180201:1101} represent the infinitesimals for the independent variables $y^4=x^{\prime 1}$ and $y^5=x^{\prime 2}$, respectively.

With the ansatz \eqref{180201:1057}-\eqref{180201:1125} and the result \eqref{180201:0924}-\eqref{180201:0930} for $\mathsf{E}_1$, a combined symmetry analysis for $\mathsf{E}_1$-$\mathsf{E}_3$ inevitably leads to the relations (Eqs.$\,$[40-45])
\begin{align}
d^1_1(\vx)&= 2c^{11}(\vx)-c^{11}_1(\vx)x^1-c^{12}_1(\vx)x^2,\label{180201:1425}\\[0.5em]
d^1_2(\vx)&= -c^{11}_2(\vx)x^1-c^{12}_2(\vx)x^2,\\[0.5em]
d^2_1(\vx)&= c^{12}_1(\vx)x^1-c^{11}_1(\vx)x^2,\\[0.5em]
d^2_2(\vx)&= 2c^{11}(\vx)+c^{12}_2(\vx)x^1-c^{11}_2(\vx)x^2,\label{180201:1426}
\end{align}
with
\begin{equation}
3c^{11}_1=-c^{12}_2,\quad 3c^{11}_2=c^{12}_1,\;\;\text{and hence:}\;\; c^{11}_{11}+c^{11}_{22}=0,\quad  c^{12}_{11}+c^{12}_{22}=0,\label{180201:1702}
\end{equation}
and the further results (Eq.$\,$[39] and Eq.$\,$[52])\footnote{Note that $C_1+C_2$ in \cite{Grebenev17} corresponds exactly to $C(\omega)$ in \eqref{180201:0930}; see p.$\,$16 where ``the constant $C$ [in Eq.$\,$29 or A.4] was presented as a sum of the two constants $C=C_1+C_2$".}
\begin{align}
\xi^6&=2c^{11}(\vx)\omega^\prime,\label{180213:1433}\\[0.5em]
\eta_{f_2}&=-\Big(8c^{11}(\vx)+C(\omega)\Big)f_2+b^{\prime}(t,\vy),\label{180203:1632}
\end{align}
where $b^\prime$ is an arbitrary solution to the equations $\mathsf{E}_2$ and $\mathsf{E}_3$ in correspondence to the two arbitrary solutions $b^1$ and $b^2$ given in \eqref{180201:1224} for $\mathsf{E}_1$, i.e.,
\begin{equation}
\left.
\begin{aligned}
b^1(t,\vx,\omega)+\frac{1}{2\pi}\int d^2\vx^\prime d\omega^\prime \omega^\prime\frac{x^2-x^{\prime 2}}{|\vx-\vx^\prime|^2}b^\prime(t,\vy)=0,\\[0.5em]
b^2(t,\vx,\omega)-\frac{1}{2\pi}\int d^2\vx^\prime d\omega^\prime \omega^\prime\frac{x^1-x^{\prime 1}}{|\vx-\vx^\prime|^2}b^\prime(t,\vy)=0.
\end{aligned}
~~~~~\right\}
\end{equation}
To note is that the dependencies of the infinitesimals $\xi^3$, $\eta^0$, $\eta^1$ and $\eta^2$ in \eqref{180201:0924}-\eqref{180201:1224} stay unchanged after this extended analysis, i.e., augmenting the symmetry analysis for $\mathsf{E}_1$ by including $\mathsf{E}_2$ and $\mathsf{E}_3$ does not restrict these infinitesimals any further; they simply are not effected and thus\pagebreak[4] remain unchanged by this extension. Employing the specific ansatz \eqref{180201:1057}-\eqref{180201:1125} and the result \eqref{180201:1425}-\eqref{180201:1426}, the latter three infinitesimals can at least be explicitly written out as (Eq.$\,$[51])
\begin{equation}
\left.
\begin{aligned}
\eta^0&=-\Big(6c^{11}(\vx)+C(\omega)\Big)f_1+b^0(t,\vx,\omega),\\[0.5em]
\eta^1&=-\Big(3c^{11}(\vx)+C(\omega)\Big)J^1+c^{12}(\vx)J^2+b^1(t,\vx,\omega),\\[0.5em]
\eta^2&=-c^{12}(\vx)J^1-\Big(3c^{11}(\vx)+C(\omega)\Big)J^2+b^2(t,\vx,\omega).\label{180201:1521}
\end{aligned}
~~~~~\right\}
\end{equation}
Now, by including also the last two remaining equations into the symmetry analysis, namely the two consistency conditions $\mathsf{E}_4$ \eqref{180201:1006} and  $\mathsf{E}_5$ \eqref{180201:1007}, further restrictions for the infinitesimals can be expected. Including first the latter equation $\mathsf{E}_5$, we just obtain the trivial restriction (Eq.$\,$[53])
\begin{equation}
b^0(t,\vx,\omega)=\int d\omega^\prime b^\prime(t,\vy),\label{180201:1523}
\end{equation}
meaning that equation $\mathsf{E}_5$ \eqref{180201:1007} already transforms as an invariant under the generating transformations \eqref{180201:1057}-\eqref{180201:1521}. In other words, when taking along the constraint \eqref{180201:1523}, equation $\mathsf{E}_5$ is fully compatible to the already determined symmetries of subsystem $\mathsf{E}_1$-$\mathsf{E}_3$; no symmetries are broken when augmenting this system by $\mathsf{E}_5$.

When including equation $\mathsf{E}_4$ \eqref{180201:1006}, however, the situation is different: Besides the trivial restriction (Eq.$\,$[53])\footnote{Note that the explicit form of the restrictions \eqref{180201:1600} and \eqref{180201:1601} can also be represented differently, for example, when splitting the arbitrary function $C(\omega)$ additively into two separate ones $C(\omega)=C_1(\omega)+C_2(\omega)$, as has been done in \cite{Grebenev17}. For instance, if $C_2$ is linked to $b^0$ and $C_1$ to $\xi^3$, then \eqref{180201:1600} and \eqref{180201:1601} can also be equivalently written as $\int\! d\omega b^0(t,\vx,\omega)=\int\! d\omega \hspace{0.25mm}C_2(\omega)$ and $\xi^3_3=6c^{11}(\vx)+C_1(\omega)$, respectively.}
\begin{equation}
\int d\omega b^0(t,\vx,\omega)=0,\label{180201:1600}
\end{equation}
we also obtain the crucial restriction\footnote{Restriction \eqref{180201:1601} guarantees that the for the combined system $\mathsf{E}_1$-$\mathsf{E}_5$ determined symmetry transformation is universally valid for {\it all} possible solutions of the 1-point PDF $f_1$, which essentially is also the purpose of every symmetry analysis: To find transformations that leave equations invariant independent of the particular structure they may give as solutions for the dependent variables --- see Sec.$\,$\ref{Sec.2.2} for the derivation of \eqref{180201:1601} and for a more detailed discussion on that issue.} (see Sec.$\,$\ref{Sec.2.2} for a detailed proof)
\begin{equation}
\xi^3_3=6c^{11}(\vx)+C(\omega),\label{180201:1601}
\end{equation}
which forces the function $c^{11}$ to be a constant now not depending on the spatial coordinate~$\vx$, simply because the left-hand side $\xi^3_3$ is, according to result \eqref{180201:0924}, a function of $\omega$ only: $\xi^3=\xi^3(\omega)$. Hence, since this result is globally valid for {\it all} $\omega\in\mathbb{R}$, {\it including} the case $\omega=0$, the above restriction \eqref{180201:1601} is equivalent to the combined restriction\footnote{Note that according to result \eqref{180201:1702}, the constraint \eqref{180201:1703} also restricts its dual function $c^{12}(\vx)$ to be a constant: $c^{12}_1(\vx)=0$ and $c^{12}_2(\vx)=0$.}
\begin{equation}
c^{11}_1(\vx)=0\quad\text{and}\quad c^{11}_2(\vx)=0,\quad \forall\omega\in\mathbb{R}, \label{180201:1703}
\end{equation}
which contradicts the ansatz made for $\xi^3$ (Eq.$\,$[38]) in \cite{Grebenev17}, where, also for the particular case of zero vorticity $\omega=0$, this function is wrongly and misleadingly prescribed to be non-constant: $c^{11}_1(\vx)\neq 0$ and $c^{11}_2(\vx)\neq0$.

Hence, due to this constraint \eqref{180201:1601}, or equivalently due to \eqref{180201:1703}, the local conformal invariance for $\mathsf{E}_1$-$\mathsf{E}_5$ \eqref{180131:0807}-\eqref{180201:1007}, which itself as a combined system represents the first equation in the infinite and unclosed hierarchy of LMN vorticity equations, cannot be confirmed as claimed in \cite{Grebenev17}, also not for the particular case $\omega=0$.

Even if this proof is already fully sufficient to show that \cite{Grebenev17} has been refuted, we will nevertheless examine their false and thus misleading conclusion in more detail in the next section, by looking at it from different perspectives in order to gain a better understanding of all the processes involved.

\section{Revealing and correcting the mistake in \cite{Grebenev17}}

\subsection{First perspective: The smoothness-axiom of Lie-groups}

The heart of their non-correctible error in \cite{Grebenev17} lies in the interplay between result Eq.$\,$[38] and Eq.$\,$[30]:
\begin{align}
&\xi^3=\big[6c^{11}(\vx)+C_1\big]\omega,\label{200420:1411}\\[0.5em]
&\xi^3_1=\xi^3_2=0.\label{200420:1412}
\end{align}
The solution constraint \eqref{200420:1412} says that the infinitesimal~$\xi^3$, given by \eqref{200420:1411}, should not~depend on the spatial coordinates $x^1$ and $x^2$. Sure, at first glance \eqref{200420:1411} and \eqref{200420:1412} stand in conflict with each other, because to force a persistent spatial dependence with \eqref{200420:1411} is obviously not compatible with the spatial independence
as demanded by \eqref{200420:1412}. But at a second glance, when particularly looking at the functional structure of \eqref{200420:1411}, it seems that this conflict can be easily resolved if $\omega=0$ is chosen, as the
authors then did after Eq.$\,$[46].

Because now, with this specification $\omega=0$, result \eqref{200420:1411} turns to $\xi^3=0$ with which \eqref{200420:1412} then turns into $0=0$ and which therefore, according to the rationale of \cite{Grebenev17}, can be successfully removed from the invariance group simply because this constraint \eqref{200420:1412} gets identically satisfied when evaluated at $\omega=0$. That the constraint \eqref{200420:1412}, i.e. Eq.$\,$[30], is indeed removed from the invariance group can be explicitly seen, e.g., in their incorrect final result~Eq.$\,$[61] and its consequence Eq.$\,$[64], in that both these results do not come along with the constraint~Eq.$\,$[30] anymore, or, in the flawed\footnote{An independent proof for this claim will be given in the next section, Sec.$\,$\ref{Sec.2.2}.} and misleading invariance proof of the probability measure $\mu$ in Sec.$\,$3.3, which explicitly shows that Eq.$\,$[30] is not part of their proof anymore.

It's clear what the implications are when incorrectly removing constraint \eqref{200420:1412} from the invariance group for $\omega=0$: The function $c^{11}$ in \eqref{200420:1411} need not to be reduced to a global constant since it need not to comply with \eqref{200420:1412} anymore (due to its ``non-restricting" form $0=0$),\linebreak[4] but can remain to be a general function on the spatial coordinates $x^1$ and $x^2$, which then, along with the conditions Eqs.$\,$[44-45], allows for the desired conformal invariance Eqs.$\,$[34-37] not to get broken. But this reasoning is both wrong and seriously misleading as we will prove next.

\subsubsection{Proof that constraint \eqref{200420:1412} may not be removed from invariance group even if $\omega=0$}
Even when putting $\omega=0$ in order to enforce compatibility between \eqref{200420:1412} and \eqref{200420:1411}, and the constraint \eqref{200420:1412} itself pretends to be in the non-restrictive form $0=0$, the authors do not have a magic wand to simply let \eqref{200420:1412} disappear from the invariance group. The constraint \eqref{200420:1412} is still there and active even after putting $\omega=0$, simply because \eqref{200420:1412} is permanently valid for {\it all}~real~numbers of $\omega$, and that without any restrictions, which eventually is a crucial information not explicitly mentioned by the authors. But what does this additional information imply now? Well, since~\eqref{200420:1412} is {\it continuously} valid for {\it all} $\omega\in\mathbb{R}$ without any restrictions (which we will discuss at length further below and prove in detail in Appendix \ref{Sec.A}),
we can take, for example, any differential consequence of \eqref{200420:1412} according to this variable without restrictions, for which we will then get further combined constraint equations also {\it continuously} valid for {\it all} $\omega\in\mathbb{R}$, on account of the underlying smoothness-axiom of Lie-groups. For example, due to the global existence~of~\eqref{200420:1412}
\begin{equation}
\frac{\partial\xi^3}{\partial x^1}=\frac{\partial\xi^3}{\partial x^2}=0, \;\;\forall \omega\in\mathbb{R},
\label{190713:1927}
\end{equation}
we thus can imply the following differential consequence:
\begin{equation}
\frac{\partial^2 \xi^3}{\partial\omega\partial x^1}=\frac{\partial^2 \xi^3}{\partial\omega\partial x^2}=0, \;\;\forall \omega\in\mathbb{R}
\quad\;\;\Longleftrightarrow \quad\;\;
\frac{\partial}{\partial x^1}\left(\frac{\partial \xi^3}{\partial \omega}\right)=\frac{\partial}{\partial x^2}\left(\frac{\partial \xi^3}{\partial \omega}\right)=0, \;\;\forall \omega\in\mathbb{R}.
\label{190712:1446}
\end{equation}
The right-hand side of this implication \eqref{190712:1446} tells us now that the function $\xi_3^3:=\partial_\omega \xi^3$ should not depend on the spatial coordinates $x^1$ and $x^2$ for {\it any} value of $\omega$ as well, i.e.,
\begin{equation}
\frac{\partial\xi_3^3}{\partial x^1}=\frac{\partial\xi^3_3}{\partial x^2}=0, \;\;\forall \omega\in\mathbb{R},
\label{190712:1626}
\end{equation}
where initially in \eqref{190712:1446}, and this is important, we explicitly made use of the smoothness-axiom of Lie-groups which allows for the interchanging of partial derivatives on its elements.

Hence, besides $\xi^3$, also $\xi^3_3$ should be spatially independent for  {\it any} value of $\omega$, including $\omega=0$.\linebreak[4] But this result causes a problem now. While the space-dependent result \eqref{200420:1411}
\begin{equation}
\xi^3=\big[6c^{11}(\vx)+C_1\big]\omega,
\label{200421:0931}
\end{equation}
could still be made compatible with constraint \eqref{190713:1927} by choosing $\omega=0$, this clearly does not work anymore for the next higher-order constraint \eqref{190712:1626}, since $\xi^3$ is linear in $\omega$.
Inserting \eqref{200421:0931} into \eqref{190712:1626} then leads to
\begin{equation}
\frac{\partial c^{11}(\vx)}{\partial x^1}=\frac{\partial c^{11}(\vx)}{\partial x^2}=0, \;\;\forall \omega\in\mathbb{R},
\label{190713:2318}
\end{equation}
which will reduce the spatial function $c^{11}$ to a global constant and thus, as a final result, the desired conformal invariance is broken, in particular also for $\omega=0$.\qed

It's clear that the crucial aspect to obtain the above result \eqref{190712:1626} is that \eqref{200420:1412} has to be valid for {\it all} $\omega\in\mathbb{R}$, as explicitly and transparently written in \eqref{190713:1927}:
\begin{equation*}
\frac{\partial\xi^3}{\partial x^1}=\frac{\partial\xi^3}{\partial x^2}=0, \;\;\forall \omega\in\mathbb{R}.
\label{200421:1119}
\end{equation*}
Indeed, this unrestricted condition on constraint \eqref{200420:1412}, namely that $\xi^3$ is globally independent on the spatial coordinates for any $\omega$, we already have given the proof through result \eqref{180201:0924}-\eqref{180201:0930},\linebreak[4] a result that is shared\footnote{Up to the concern stated in footnote no.\ref{fn1}, p.$\,$\pageref{fn1}.} by \cite{Grebenev17}, but in its consequence has not been interpreted correctly by them. The correct interpretation of result \eqref{180201:0924} is that $\xi^3$ shows no other dependence than solely on $\omega$ and that this dependence is unrestricted, i.e., it is not constrained by any hidden or explicit condition on $\omega$:
\begin{equation}
\xi^3\equiv \xi^3(\omega),\: \text{unrestrictedly for~{\it all}~$\omega\in\mathbb{R}$}.
\end{equation}
In other words, when performing a thorough symmetry investigation, as we did to obtain the correct result \eqref{180201:0924}-\eqref{180201:0930}, it shows that the constraint Eq.$\,$[30] as given in \cite{Grebenev17} is not complete. It correctly has to be extended to:
\begin{equation}
\xi^3_1=\xi^3_2=\xi^3_4=\xi^3_5=\xi^3_6=\xi^3_0=0, \;\;\forall \omega\in\mathbb{R}, \;\;\text{while}\;\; \xi^3_3\neq 0.
\label{190712:1711}
\end{equation}
To avoid any misunderstandings of this result, the following should be~noted: Constraint~\eqref{190712:1711} is {\it not} an assumption, but the result of a thorough symmetry investigation, which we carefully checked both by hand as well as by using third-party computer software\footnote{In Appendix \ref{Sec.A} we provide an explicit proof of \eqref{190712:1711}, in that we perform a complete invariance analysis of the defining local equation $\mathsf{E}_1$~\eqref{180131:0807} that results to \eqref{190712:1711}.} that systematically calculates all Lie group symmetries of differential equations automatically. Because \eqref{190712:1711} is essentially nothing else than the full symmetry solution for the infinitesimal $\xi^3$ of the local (differential) part of the considered system, which in \cite{Grebenev17} is given by Eq.$\,$[6] and herein by~\eqref{180131:0807}. But \cite{Grebenev17} failed to give the full symmetry solution, simply because the decisive and crucial information $\forall \omega\in\mathbb{R}$ of result~\eqref{190712:1711} is missing and thus not part of their solution as they derived it in Sec.$\,$A.1.

\subsection{Second perspective: Non-invariance of the normalization condition $\boldsymbol{\mathsf{E}_4}$\label{Sec.2.2}}

When performing a Lie-group symmetry analysis on the governing equation $\mathsf{E}_1$ \eqref{180131:0807}, it provides us with two strong results: (i) The infinitesimal $\xi^3$  \eqref{180201:0924} is only a function of its defining vorticity variable: $\xi^3=\xi^3(\omega)$, and (ii) the infinitesimals $\eta^i$ \eqref{180201:1224} for the dependent variables can be~supplemented by a function $C$ \eqref{180201:0930}, depending also only on the vorticity variable: $C=C(\omega)$.\linebreak[4] In particular, when augmenting the symmetry analysis by also including the remaining equations $\mathsf{E}_2$-$\mathsf{E}_5$, this twofold result from $\mathsf{E}_1$ is of global nature, meaning that $\xi^3$ and $C$ should and may not depend on the spatial variable $\vx$ for {\it all} $\omega\in\mathbb{R}$, including also the zero vorticity case $\omega=0$.

In \cite{Grebenev17}, however, the following unexplained and misleading ansatz for $\xi^3$ is made (Eq.$\,$[38]):
\begin{equation}
\xi^3=\Big(6c^{11}(\vx)+C_1\Big)\omega,\;\;\text{for}\;\; c^{11}_1(\vx)\neq 0\;\;\text{and}\;\; c^{11}_2(\vx)\neq 0,\label{180201:2127}
\end{equation}
which obviously, as explained above, is not compatible with the symmetry result as stipulated by the governing equation $\mathsf{E}_1$ \eqref{180131:0807}. The argument in \cite{Grebenev17}, however, is that for the specific case $\omega=0$ this conflict is resolved. Although this argument itself is correct, since a zero infinitesimal $\xi^3=0$ is inherently independent of any variables whatsoever, they overlooked the fact that their {\it local} condition, which just only holds for the single value $\omega=0$, cannot be employed to transform the normalization condition $\mathsf{E}_4$ \eqref{180201:1006}, which obviously constitutes a {\it non-local} constraint equation.\footnote{The same mistake also has been made in Sec.$\,$3.3 in \cite{Grebenev17}, which, if corrected, invalidates their claim that the probability measure $\mu(t,\vx,\omega)=f_1(t,\vx,\omega)d\omega$ is local-conformally invariant. On the one side their mistake is that Eq.$\,$[38] for $\omega\neq 0$ {\it may not} be used to transform $\mu$ since it is inconsistent to the symmetry transform of the governing Eq.$\,$[6], and on the other side their mistake is that Eq.$\,$[38] for $\omega=0$ {\it cannot} be used to transform~$\mu$ since $\omega$ is then rigidly fixed and thus not variable anymore.} The reason is that $\mathsf{E}_4$ is a global relation that sums over {\it all} vorticity values $\omega\in\mathbb{R}$, and not only locally for $\omega=0$. To~therefore correctly transform this global constraint $\mathsf{E}_4$
\begin{equation}
\int d\omega f_1=1,\label{180201:2145}
\end{equation}
where $\omega$ needs to be varied by $d\omega$ over the whole (infinite) integration range, one has to use a transformation rule for $\omega$ that is globally valid for all values, and not only for the fixed value~$\omega=0$. Hence, to invariantly transform \eqref{180201:2145} in line with~$\mathsf{E}_1$, the transformation rule~\eqref{180201:2127} is not the correct choice, since it is only valid for $\omega=0$ and thus, as a fixed single value, cannot be varied in $\omega$, while for $\omega\neq 0$, as already said, rule \eqref{180201:2127} has to be discarded, simply because it is incompatible to the existing constraint $\xi^3_1=\xi^3_2=0$, that means incompatible to the general result $\xi^3=\xi^3(\omega)$ for~$\mathsf{E}_1$ to be invariant.

Hence it is clear that only the following globally valid ansatz (up to order $\mathcal{O}(\epsilon^2)$ in the group parameter $\epsilon$)
\begin{equation}
\tilde{\omega}=\omega+\epsilon\cdot \xi^3(\omega),\quad\forall\omega\in\mathbb{R},\label{180201:2313}
\end{equation}
will invariantly transform the global constraint $\mathsf{E}_4$ \eqref{180201:2145} in line with $\mathsf{E}_1$. The associated infinitesimal constraint that will be induced as result then has the form
\begin{align}
0 &= 1-\int d\tilde{\omega}\tilde{f}_1=1-\int d\omega\left|\frac{\partial\tilde{\omega}}{\partial\omega}\right|\big(f_1+\epsilon \eta^0+\mathcal{O}(\epsilon^2)\big)
\underset{\eqref{180201:2313}}{=}1-\int d\omega \big|1+\epsilon\xi^3_3\big|\big(f_1+\epsilon \eta^0\big)+\mathcal{O}(\epsilon^2)\nonumber\\[0.5em]
&\underset{\epsilon\ll 1}{=}1-\int d\omega \big(1+\epsilon\xi^3_3\big)\big(f_1+\epsilon \eta^0\big)+\mathcal{O}(\epsilon^2)
=1-\int d\omega \big(f_1+\epsilon\big(\eta^0+\xi^3_3f_1\big)\big) +\mathcal{O}(\epsilon^2)\nonumber\\[0.5em]
&\underset{\eqref{180201:1521}}{=}
1-\int d\omega \big(f_1+\epsilon\big(-6c^{11}f_1-Cf_1+b^0+\xi^3_3f_1\big)\big) +\mathcal{O}(\epsilon^2)\nonumber\\[0.5em]
&\underset{\eqref{180201:1600}\,\&\,\eqref{180201:2145}}{=}\int d\omega\big(6c^{11}+C-\xi^3_3\big)f_1 +\mathcal{O}(\epsilon),\label{180202:1946}
\end{align}
which, if we seek for a symmetry transformation that is valid for {\it all} possible solutions $f_1$, is equivalent to the constraint \eqref{180201:1601}
\begin{equation}
6c^{11}(\vx)+C(\omega)-\xi^3_3(\omega)=0,\label{180202:1947}
\end{equation}
that now forces $c^{11}$ to be a constant,\footnote{Another independent but equivalent argument that $c^{11}$ needs to be a constant is to recognize that the non-local determining equation \eqref{180202:1946} is living in a jet space  where $\vx$ and $f_1$ are jet coordinates defined by the underlying symmetry analysis of the governing system \eqref{180130:2115}-\eqref{180130:2158}, i.e., a particular designed space where $\vx$ and $f_1$ are defined to each other as independents: $\partial_{x^1}f_1=\partial_{x^2}f_1=0$. Then, by re-writing \eqref{180202:1946} as $6c^{11}+\int d\omega(C-\xi^3_3)f_1=0$ (since $c^{11}$ by construction is independent of $\omega$) and by taking the spatial derivatives on both sides, one directly obtains the overall consistent result $c^{11}_1=c^{11}_2=0$, simply due to that $C$ and $\xi^3$ on the one side are independent functions of~$\vx$ and on the other that $f_1$ is a jet variable with respect to $\vx$. --- On jet spaces in general, see e.g. \cite{Olver93}.} due to that $\xi^3$, according to the rule \eqref{180201:2313}, is only a function of $\omega$ not depending on $\vx$.\qed

As exercised in \cite{Grebenev17}, which again refers to \cite{Ibragimov02}, the determining equation \eqref{180202:1947} can also be derived alternatively by noting that the non-local determining symmetry equation \eqref{180202:1946} can be split with respect to group variable $f_1$ using variational differentiation. Since the bracketed term in \eqref{180202:1946} does not depend on $f_1$, taking the variational or functional derivative $(\delta/\delta f_1(\bar{\omega}))$ of this equation then leads to the same local result \eqref{180202:1947}:
\begin{align}
0&=\frac{\delta}{\delta f_1(\bar{\omega})}\int d\omega\big(6c^{11}+C-\xi^3_3\big)f_1\nonumber\\[0.5em]
&=\int d\omega \big(6c^{11}+C-\xi^3_3\big)\delta(\omega-\bar{\omega})=6c^{11}(\vx)+C\big|_{\omega=\bar{\omega}}-\xi^3_3\big|_{\omega=\bar{\omega}}.
\end{align}
Note here that it is valid to take the variational derivative of equation \eqref{180202:1946} since $f_1$ can be {\it continuously} varied to still satisfy equation \eqref{180202:1946} by just choosing the non-constant coefficient or pre-factor of $f_1$ appropriately, trivially of course as \eqref{180202:1947}. In contrast of course to the defining equation \eqref{180201:2145} itself, which determines or fixes $f_1$ and which thus cannot be continuously varied: Any arbitrary non-zero functional variation of $f_1$ will violate the constraint \eqref{180201:2145}, as can be clearly seen by taking the variational derivative $(\delta/\delta f_1(\bar{\omega}))$ of this constraint
\begin{equation}
0=\frac{\delta}{\delta f_1(\bar{\omega})}\Big(-1+\int d\omega f_1\Big)=\int d\omega \delta(\omega-\bar{\omega})=1,\label{180202:2329}
\end{equation}
turning the constraint \eqref{180201:2145} thus into the contradiction $1=0$. This conflict just tells us that equation \eqref{180201:2145} cannot be functionally varied simply because it defines and determines the function $f_1$, similar as in the usual variation for real numbers if we would fix a variable to a certain value, say $x=1$, then any variation on it would be meaningless, since $x$ is defined or determined strictly as $1$. Evidently, taking the variation of $x=1$ leads to the same conflict
\begin{equation}
0=\frac{\partial}{\partial x}(-1+x)=1,
\end{equation}
as in \eqref{180202:2329} for the functional variation of the defining and determining equation \eqref{180201:2145} for $f_1$.

Back again to the general result \eqref{180202:1947} by choosing $\xi^3=0$, as it would be the case in~\cite{Grebenev17} when applying in Eq.$\,$[38] their necessary zero-vorticity constraint $\omega=0$~(Eq.$\,$[32]),\linebreak[4] we note that their results for the infinitesimals $\eta^0$ and $\eta^\prime$ as given by Eqs.$\,$[51-52] are incorrect. Instead of an unrestricted $C=C_1+C_2$, the constant restriction $C=-6c^{11}$ has to be used in their results in order to be consistent with $\xi^3=0$, exactly as it is required by \eqref{180202:1947}. Hence, for $\xi^3=0$, the correct generating infinitesimals for $f_1$ and $f_2$ are given by \eqref{180201:1521} \& \eqref{180203:1632}
\begin{equation}
\eta_{f_1}\equiv\eta^0=b^0(t,\vx,\omega),\qquad \eta_{f_2}\equiv\eta^\prime=-2c^{11}f_2+b^\prime(t,\vy),\qquad \forall \omega\in\mathbb{R},\;\; c^{11}_1=c^{11}_2=0,\label{180203:1648}
\end{equation}
and not by Eqs.$\,$[51-52] as proposed in \cite{Grebenev17}. Note, since $C=-6c^{11}$ itself is a $\omega$-independent choice, result \eqref{180203:1648} is in fact globally valid for all $\omega\in\mathbb{R}$.
Regarding Sec.$\,$3.3 in \cite{Grebenev17}, it is thus clear that for the above choice $\xi^3=0$, $\forall\omega\in\mathbb{R}$, and its resulting transform \eqref{180203:1648}, the probability measure $\mu=f_1d\omega$ (Eq.$\,$[68]) remains to be invariant, but not local-conformally anymore as claimed since $c^{11}$, according to \eqref{180203:1648}, is a spatial constant~now.

\subsection{Third perspective: The non-exceptional role of $\boldsymbol{\omega=0}$\label{Sec.2.3}}

It is clear that when analyzing any system of equations on symmetries, as for example for the case $\mathsf{E}_1$-$\mathsf{E}_5$ \eqref{180131:0807}-\eqref{180201:1007} considered herein, the symmetry result should not depend on the choice which subsystem is considered first and in which order it is being evaluated, i.e., no matter which direction in evaluation one takes, a symmetry analysis should always give exactly the same result, otherwise a consistent analysis is not guaranteed. For example, let us first consider the following approach: Before starting any analysis, we already specify the coordinate $\omega$ in the subsystem $\mathsf{E}_1$-$\mathsf{E}_3$ \& $\mathsf{E}_5$ to an arbitrary but fixed value, say $\omega=\omega^*$, where $\omega^*\in\mathbb{R}$ can be any value from real space, including the choice $\omega^*=0$. The initial system $\mathsf{E}_1$-$\mathsf{E}_5$ \eqref{180131:0807}-\eqref{180201:1007} then turns into the form
\begin{align}
\mathsf{E}^*_1\!:&\quad \frac{\partial J^{*0}}{\partial y^0}+ \frac{\partial J^{*1}}{\partial y^1}+\frac{\partial J^{*2}}{\partial y^2}=0,\label{180213:0925}\\[0.5em]
\mathsf{E}^*_2\!:&\quad J^{*1}+\frac{1}{2\pi}\int d^2\vx^\prime d\omega^\prime \omega^\prime\frac{x^2-x^{\prime 2}}{|\vx-\vx^\prime|^2}f_2^*=0,\label{180213:0926}\\[0.5em]
\mathsf{E}^*_3\!:&\quad J^{*2}-\frac{1}{2\pi}\int d^2\vx^\prime d\omega^\prime \omega^\prime\frac{x^1-x^{\prime 1}}{|\vx-\vx^\prime|^2}f_2^*=0,\label{180213:0927}\\[0.5em]
\mathsf{E}_4\!:&\quad 1-\int d\omega f_1=0,\label{180213:0928}\\[0.5em]
\mathsf{E}_5^*\!:&\quad f_1^*-\int d\omega^\prime f_2^*=0,\label{180213:0929}
\end{align}
where $(J^{*i},f_2^*)$ are the functions $(J^i,f_2)$ evaluated at $\omega=\omega^*$:
\begin{equation}
f_1^*\equiv J^{*0}=J^0\big|_{\omega=\omega^*}\equiv f_1\big|_{\omega=\omega^*},\qquad J^{*1;2}=J^{1;2}\big|_{\omega=\omega^*},\qquad f_2^{*}=f_2\big|_{\omega=\omega^*}.\label{180213:1400}
\end{equation}
Instead of 11 jet coordinates $(y^0,\vy,\vJ,f_2)$ for the initial system \eqref{180131:0807}-\eqref{180201:1007}, a symmetry analysis of \eqref{180213:0925}-\eqref{180213:0929} now defines an extended jet space with 12 coordinates $(y^0,\vy,\vJ^*,f_2^*,f_1)$, where the additional coordinate $f_1$ is related to $J^{*0}$ via \eqref{180213:1400}. Formally, the four equations $\mathsf{E}^*_1$-$\mathsf{E}^*_3$ \& $\mathsf{E}^*_5$ can now be identified as a system being independent of the jet coordinate $y^3=\omega$. Performing a symmetry analysis on this reduced subsystem for the ansatz \eqref{180201:1057}-\eqref{180201:1125}, one yields the same results
\eqref{180201:1425}-\eqref{180201:1523} as before in just replacing the $\omega$-dependent infinitesimals $(\eta^i,\eta_{f_2})$ in \eqref{180201:1521} and \eqref{180203:1632} by their corresponding $\omega$-independent infinitesimals $(\eta^{*i},\eta^*_{f^*_2})$:
\begin{equation}
\left.
\begin{aligned}
\eta^{*0}&=-\big(6c^{11}(\vx)+C^*\big)f_1^*\,\equiv\, \eta^*_{f_1^*},\\[0.5em]
\eta^{*1}&=-\big(3c^{11}(\vx)+C^*\big)J^{*1}+c^{12}(\vx)J^{*2},\\[0.5em]
\eta^{*2}&=-c^{12}(\vx)J^{*1}-\big(3c^{11}(\vx)+C^*\big)J^{*2},\\[0.5em]
\eta^*_{f^*_2}&=-\big(8c^{11}(\vx)+C^*\big)f^*_2,
\label{180213:1453}
\end{aligned}
~~~~~\right\}
\end{equation}
where $C^*$ is any arbitrary constant and where, for simplicity and convenience, the solutions\linebreak[4] $b^{*i}$ and $b^{*\prime}$ were chosen as the trivial zero solutions, simply because of not being relevant here for the present discussion on consistency. Now, in using the defining relation \eqref{180213:1400}, we can read~off the unreduced infinitesimal~$\eta^0$ for the jet coordinate $f_1$ from \eqref{180213:1453} as\footnote{$C^*$ in \eqref{180213:1453} is then defined as the evaluation $C(\omega)\big|_{\omega=\omega^*}\equiv C^*$.}
\begin{equation}
\eta^0=-\Big(6c^{11}(\vx)+C(\omega)\Big)f_1\,\equiv\, \eta_{f_1},\label{180213:1543}
\end{equation}
which is necessary now in order to find the infinitesimal $\xi^3\equiv\xi_\omega$ from the last and remaining equation $\mathsf{E}_4$~\eqref{180213:0928} in this system. Obviously, since \eqref{180213:1543} matches the result \eqref{180201:1521}, the symmetry analysis of $\mathsf{E}_4$ \eqref{180213:0928} is identical to the one performed in the previous section~\eqref{180202:1946}, with the same result \eqref{180202:1947}, however, now without any constraints on the infinitesimal $\xi^3$:
\begin{equation}
6c^{11}(\vx)+C(\omega)-\xi^3_3(t,\vx,\omega)=0,\label{180213:1606}
\end{equation}
i.e., where $\xi^3$ can freely depend now on all independent variables involved,\footnote{Note that since $\xi^3$ is associated to the 1-point quantity $f_1$ there is no dependence on any 2-point coordinates.} simply because the constructed $\omega$-independence of the newly defined local equation $\mathsf{E}^*_1$ \eqref{180213:0925} cannot give or lead to any restrictions on $\xi^3$, as it was the case before for the $\omega$-dependent  local equation $\mathsf{E}_1$~\eqref{180131:0807}.

Solving \eqref{180213:1606} explicitly for $\xi^3$, the only solution that is in accordance or in line with subsystem $\mathsf{E}^*_1$-$\mathsf{E}^*_3$ \& $\mathsf{E}^*_5$ is given by the particular solution
\begin{equation}
\xi^3(t,\vx,\omega)=\Big(6c^{11}(\vx)+C(\omega)\Big)(\omega-\omega^*)-\int_{\omega^*}^\omega  \frac{dC(\bar{\omega})}{d\bar{\omega}}(\bar{\omega}-\omega^*)\, d\bar{\omega},\label{180213:1725}
\end{equation}
since for $\omega=\omega^*$ this infinitesimal turns zero $\xi^3|_{\omega=\omega^*}=0$, with the effect then that under this transformation the value $\omega=\omega^*$ gets mapped to the same value again:
$\omega^*=\omega\mapsto\tilde{\omega}=\omega^*$, and therefore keeping subsystem $\mathsf{E}^*_1$-$\mathsf{E}^*_3$ \& $\mathsf{E}^*_5$ thus invariant.

Hence, it seems that the analysis of \cite{Grebenev17} just got generalized to arbitrary vorticity isolines, since the result \eqref{180213:1725} is not restricted to the particular value of a zero-vorticity isoline
$\omega^*=0$, as in Eq.$\,$[38] in \cite{Grebenev17}. Hence we could say that we have shown local conformal invariance for the 2D LMN vorticity equations (up to second order in the LMN chain of equations) on all its vorticity isolines, that is, for all values of $\omega\in\mathbb{R}$. But, unfortunately, that is not the case. Because, when looking again at the determining equation~\eqref{180213:1606} when evaluated at $\omega=\omega^*$ and equivalently re-written as
\begin{equation}
\xi^3_3\big|_{\omega=\omega^*}=6c^{11}(\vx)+C(\omega^*),\;\; \text{where}\;\; \xi^3_{31}\neq 0,\;\; \xi^3_{32}\neq 0,\label{180213:1824}
\end{equation}
which in this section was obtained by first putting $\omega$ to a fixed value $\omega^*$ and then by performing a symmetry analysis, this equation \eqref{180213:1824} only constitutes an overall consistent equation if the same result is also obtained when reversing this procedure: First by performing a symmetry analysis and then by specifying $\omega=\omega^*$. For this reverse direction, however, the governing equations are $\mathsf{E}_1$-$\mathsf{E}_5$ \eqref{180131:0807}-\eqref{180201:1007}, for which the corresponding result to \eqref{180213:1824} is then given by~\eqref{180202:1947}
\begin{equation}
\xi^3_3\big|_{\omega=\omega^*}=6c^{11}(\vx)+C(\omega^*),\;\; \text{where}\;\; \xi^3_{31}=\xi^3_{32}=0\label{180213:1849}.
\end{equation}
The decisive difference between these two equations \eqref{180213:1824} and \eqref{180213:1849} is that their left-hand sides show different dependencies: While the left-hand side of \eqref{180213:1824} depends in general on $\vx$, the left-hand side of \eqref{180213:1849} is strictly independent of $\vx$. Hence, in order to obtain a consistent result for $\xi^3_3$, the spatial function $c^{11}(\vx)$ has to be reduced to a constant, i.e., $c^{11}_1=c^{11}_2=0$; only then can the two equations \eqref{180213:1824} and \eqref{180213:1849} be matched. This completes the third independent proof, demonstrating again that local conformal invariance cannot be confirmed as proclaimed in \cite{Grebenev17}, de facto disproving their claim not only for the zero-vorticity isoline but also for all non-zero ones.\qed

\subsection{Fourth perspective: Validating potential solutions via integral consequences}

In this approach we assume that we have a solution for the 1-point and 2-point PDF $(f_1,f_2)$, obtained, for example, by a direct numerical simulation (DNS) of the (inviscid) deterministic Navier-Stokes equations for some specific unbounded flow configuration. Obviously, this solution $(f_1,f_2)$ will then satisfy identically its defining PDF equations $\mathsf{E}_1$-$\mathsf{E}_5$ \eqref{180131:0807}-\eqref{180201:1007}. The question now is whether this solution remains to be solution of this system $\mathsf{E}_1$-$\mathsf{E}_5$ \eqref{180131:0807}-\eqref{180201:1007} when being transformed on a zero-vorticity isoline according to the local conformal rule as proposed in \cite{Grebenev17} (Eqs.$\,$[33-45,51-52]). The answer is obtained by augmenting the defining system $\mathsf{E}_1$-$\mathsf{E}_5$ by certain integral consequences. For example, it is trivial to conclude that if the existing and available solution $(f_1,f_2)$ satisfies the differential equation $\mathsf{E}_1$ \eqref{180131:0807} along with the constraint $\mathsf{E}_4$ \eqref{180201:1006} identically, then it also satisfies identically its integral consequence
\begin{align}
0&=\int \!d\omega\left(\frac{\partial J^{0}}{\partial y^0}+ \frac{\partial J^{1}}{\partial y^1}+\frac{\partial J^{2}}{\partial y^2}\right)\nonumber\\[0.5em]
&=\frac{\partial}{\partial y^0} \underbrace{\int\! d\omega J^0}_{\underset{\eqref{180201:1006}}{=} 1}+ \int \!d\omega\left(\frac{\partial J^{1}}{\partial y^1}+\frac{\partial J^{2}}{\partial y^2}\right)
\,\equiv\, \frac{\partial}{\partial y^1}M^1+\frac{\partial}{\partial y^2}M^2,
\end{align}
where the $M^i$ are defined as: $M^i=\int\! d\omega J^i$, $i=1,2$. Hence, in the following we will consider the following augmented system
\begin{align}
\mathsf{E}^*_1\!:&\quad \frac{\partial J^{*0}}{\partial y^0}+ \frac{\partial J^{*1}}{\partial y^1}+\frac{\partial J^{*2}}{\partial y^2}=0,\label{180213:2225}\\[0.5em]
\mathsf{E}_{11}\!:&\quad \frac{\partial M^1}{\partial y^1}+ \frac{\partial M^2}{\partial y^2}=0,\label{180213:2226}\\[0.5em]
\mathsf{E}_{12}\!:&\quad M^1 -\int d\omega J^1=0,\label{180213:2227}\\[0.5em]
\mathsf{E}_{13}\!:&\quad M^2 -\int d\omega J^2=0,\label{180213:2228}\\[0.5em]
\mathsf{E}^*_2\!:&\quad J^{*1}+\frac{1}{2\pi}\int d^2\vx^\prime d\omega^\prime \omega^\prime\frac{x^2-x^{\prime 2}}{|\vx-\vx^\prime|^2}f_2^*=0,\label{180213:2229}\\[0.5em]
\mathsf{E}^*_3\!:&\quad J^{*2}-\frac{1}{2\pi}\int d^2\vx^\prime d\omega^\prime \omega^\prime\frac{x^1-x^{\prime 1}}{|\vx-\vx^\prime|^2}f_2^*=0,\label{180213:2230}\\[0.5em]
\mathsf{E}_4\!:&\quad 1-\int d\omega f_1=0,\label{180213:2231}\\[0.5em]
\mathsf{E}^*_5\!:&\quad f^*_1-\int d\omega^\prime f^*_2=0,\label{180213:2232}
\end{align}
which consistently extends the initial system of equations $\mathsf{E}_1$-$\mathsf{E}_5$ \eqref{180131:0807}-\eqref{180201:1007} without changing the associated solution space. Note that we here proceed as in the previous section (viz.~the third perspective), where already before the upcoming symmetry analysis the above subsystem ($\mathsf{E}_1,\mathsf{E}_2,\mathsf{E}_3,\mathsf{E}_5$) is formally reduced to the $\omega$-independent subsystem ($\mathsf{E}^*_1,\mathsf{E}^*_2,\mathsf{E}^*_3,\mathsf{E}^*_5$) in that the coordinate~$\omega$ got again specified to some arbitrary but fixed value $\omega=\omega^*$, where $\omega^*\in\mathbb{R}$, including thus also again the zero-value choice $\omega^*=0$ as in \cite{Grebenev17}.

Now, knowing that a symmetry analysis of this reduced system ($\mathsf{E}^*_1,\mathsf{E}^*_2,\mathsf{E}^*_3,\mathsf{E}^*_5$) results to~\eqref{180213:1453}, we can read off again from the underlying consistency relation \eqref{180213:1400} the corresponding unreduced infinitesimals as\footnote{It is obvious that \eqref{180214:0944} matches again the result \eqref{180201:1521}. For simplicity and convenience, the solutions $b^{i}$ and~$b^{\prime}$ were chosen as the trivial zero solutions, simply because they are not relevant for the discussion to be demonstrated herein.}
\begin{equation}
\left.
\begin{aligned}
\eta^0&=-\Big(6c^{11}(\vx)+C(\omega)\Big)f_1,\\[0.5em]
\eta^1&=-\Big(3c^{11}(\vx)+C(\omega)\Big)J^1+c^{12}(\vx)J^2,\\[0.5em]
\eta^2&=-c^{12}(\vx)J^1-\Big(3c^{11}(\vx)+C(\omega)\Big)J^2,\label{180214:0944}
\end{aligned}
~~~~~\right\}
\end{equation}

\newgeometry{left=2.50cm,right=2.50cm,top=2.50cm,bottom=2.20cm,headsep=1em}

\noindent which are necessary now to determine the infinitesimals $\xi^3$ and $\eta_{M^i}$ from the four remaining equations $\mathsf{E}_{11}$-$\mathsf{E}_{13}$~\eqref{180213:2226}-\eqref{180213:2228} and $\mathsf{E}_4$ \eqref{180213:2231}. Demanding their invariance, one obtains the following determining equations for $\eta_{M^1}$, $\eta_{M^2}$ and $\xi^3$:
\begin{equation}
\left.
\begin{aligned}
&\frac{\partial \eta_{M^1}}{\partial y^1}+\frac{\partial \eta_{M^2}}{\partial y^2}=0,\qquad\:
\frac{\partial \eta_{M^1}}{\partial M^1}-\frac{\partial \xi^1}{\partial y^1}-\frac{\partial \eta_{M^2}}{\partial M^2}+\frac{\partial \xi^2}{\partial y^2}=0,\\[0.5em]
&\frac{\partial \eta_{M^1}}{\partial M^2}-\frac{\partial \xi^1}{\partial y^2}=0,\quad\;\;
\frac{\partial \eta_{M^2}}{\partial M^1}-\frac{\partial \xi^2}{\partial y^1}=0,\quad\;\;\frac{\partial \xi^1}{\partial M^1}+\frac{\partial \xi^2}{\partial M^2}=0,\label{180214:1148}
\end{aligned}
~~~~~\right\}
\end{equation}
\begin{equation}
\eta_{M^1}-\int \!d\omega(\eta^1+\xi^3_3J^1)=0,\quad\;\; \eta_{M^2}-\int \!d\omega(\eta^2+\xi^3_3J^2)=0,\quad\;\;
\int \!d\omega(\eta^0+\xi^3_3J^0)=0,\label{180214:1150}
\end{equation}
where \eqref{180214:1148} results from equation $\mathsf{E}_{11}$ \eqref{180213:2226}, and \eqref{180214:1150} from $\mathsf{E}_{12}$ \eqref{180213:2227}, $\mathsf{E}_{13}$ \eqref{180213:2228} and
$\mathsf{E}_4$~\eqref{180213:2231}, respectively. In line with the already obtained symmetry result \eqref{180214:0944} of subsystem ($\mathsf{E}^*_1,\mathsf{E}^*_2,\mathsf{E}^*_3,\mathsf{E}^*_5$), the last equation in \eqref{180214:1150} induces the already well-known relation \eqref{180213:1606}
\begin{equation}
\xi^3_3=6c^{11}(\vx)+C(\omega),\;\; \text{where}\;\; \xi^3_{31}\neq 0,\;\; \xi^3_{32}\neq 0,\label{180214:1753}
\end{equation}
which, when inserted along with \eqref{180214:0944} into the two former equations of \eqref{180214:1150}, then gives the solution for the infinitesimals $\eta_{M^i}$ explicitly as:
\begin{align}
\eta_{M^1}&=\int \!d\omega(\eta^1+\xi^3_3J^1)\nonumber\\[0.5em]
&=\int \!d\omega\Big(\! -\big(3c^{11}(\vx)+C(\omega)\big)J^1+c^{12}(\vx)J^2 +\big(6c^{11}(\vx)+C(\omega)\big)J^1\Big)\nonumber\\[0.5em]
&=\int \!d\omega\Big( 3c^{11}(\vx)J^1+c^{12}(\vx)J^2\Big)\underset{\eqref{180213:2227}\, \&\, \eqref{180213:2228}}{=}\,  3c^{11}(\vx)M^1+c^{12}(\vx)M^2,\hspace{0.6cm}
\label{180216:1556}
\end{align}
\vspace*{-1.0em}
\begin{align}
\eta_{M^2}&=\int \!d\omega(\eta^2+\xi^3_3J^2)\nonumber\\[0.5em]
&=\int \!d\omega\Big(\! -c^{12}(\vx)J^1-\big(3c^{11}(\vx)+C(\omega)\big)J^2 +\big(6c^{11}(\vx)+C(\omega)\big)J^2\Big)\nonumber\\[0.5em]
&=\int \!d\omega\Big(\! -c^{12}(\vx)J^1+3c^{11}(\vx)J^2\Big)\underset{\eqref{180213:2227}\, \&\, \eqref{180213:2228}}{=}  -c^{12}(\vx)M^1+3c^{11}(\vx)M^2.
\label{180216:1557}
\end{align}
However, this result is inconsistent to the determining equations given by \eqref{180214:1148}, as can be seen by evaluating already the first equation\footnote{In fact, it is only the first equation in \eqref{180214:1148} that is inconsistent. The four other equations evaluate identically to zero for the considered ansatz \eqref{180201:1057}-\eqref{180201:1702}.}
\begin{align}
0&=\frac{\partial \eta_{M^1}}{\partial y^1}+\frac{\partial \eta_{M^2}}{\partial y^2}\nonumber\\[0.5em]
&=3c^{11}_1M^1+c^{12}_1M^2-c^{12}_2M^1+3c^{11}_2M^2\underset{\eqref{180201:1702}}{=} 6c^{11}_1M^1+6c^{11}_2M^2\, \neq \, 0.
\end{align}
The above relation can only be made consistent if the spatial function $c^{11}(\vx)$ is reduced to a global constant, i.e., if $c^{11}_1=c^{11}_2=0$, thus eventually breaking the local conformal invariance of the subsystem ($\mathsf{E}^*_1,\mathsf{E}^*_2,\mathsf{E}^*_3,\mathsf{E}^*_5$), where this breaking occurs not only for the specific value $\omega^*=0$, but again for all real values~$\omega^*\in\mathbb{R}$.

This result finally also answers our question stated in the beginning, namely whether the local conformal transformation as proposed in \cite{Grebenev17} will map a given solution on a zero-vorticity isoline to a new solution. The answer is clearly no,\footnote{In particular, if the set of functions $M^i=\int d\omega J^i$ is a solution, then the local-conformally transformed set $\tilde{M}^i=\int d\tilde{\omega}\tilde{J}^i$ by \cite{Grebenev17} is not a solution anymore, since the governing equation $\mathsf{E}_{11}$~\eqref{180213:2226} does not stay invariant under this transformation; only for the reduced case $c^{11}_1=c^{11}_2=0$ it will stay invariant.} simply because this particularly considered transformation constitutes no invariant transformation of the LMN vorticity equations; only the reduced case $c^{11}_1=c^{11}_2=0$ constitutes one. This completes the fourth and final independent proof that refutes the key claims made by \cite{Grebenev17}.\qed

\restoregeometry

\newgeometry{left=2.50cm,right=2.50cm,top=2.50cm,bottom=1.50cm,headsep=1em}

\section{Final remarks}

{\bf R1.} On p.$\,$8 in \cite{Grebenev17} the following is said: {\it ``It is interesting to note that (32) and (44) can be derived substituting the forms (33)-(39) into the infinitesimal form of equation~(6)."} This statement is incorrect and misleading since Eq.$\,$[32], i.e., $\omega=0$, is an external constraint that cannot be derived. Indeed, in their subsequent proof on p.$\,$9, the existence of Eq.$\,$[32] is not derived but instead used again as an independent and exogenous condition.

\noindent{\bf R2.} In Sec.$\,$\ref{Sec.2.2} we have consistently proven that the first normalization constraint of Eq.$\,$[4] in \cite{Grebenev17} in effect breaks their proposed conformal invariance, thus refuting their claim that the conclusion of a conformally invariant measure  {\it ``results from the presently conducted Lie group analysis (see appendix) that the first equation from (4) is also invariant"~[p.$\,$12].} In fact, their analysis is flawed and does not allow for such a conclusion.

To note here is that it's not a minor issue that the normalization condition Eq.$\,$[4] breaks the considered conformal invariance, e.g., by saying then let's ignore the normalization condition from the system in order to restore this invariance. The normalization condition cannot be ignored, because it's an internal condition that permanently guarantees that any PDF solution $f_n$ stays physically meaningful during its evolution. In other words, if an invariance operator for the PDF system, as given in \cite{Grebenev17} by Eq.$\,$[3], is not compatible to the normalization condition Eq.$\,$[4], then physical solutions can get mapped to unphysical ones. Therefore the normalization is an important ingredient in any PDF system and should be respected within a symmetry analysis. The same is true for all other internal constraints that go along with such a PDF system, all necessary to ensure physical PDF solutions. In this regard, please see our comment \cite{Frewer17}, which criticizes an earlier publication by V.~Grebenev \citep{Waclawczyk17}, in that new invariance groups get proposed therein which obviously are not compatible to the full PDF system when including all internal constraints, i.e., ultimately, in \cite{Waclawczyk17} non-physical symmetries are getting proposed.

\noindent{\bf R3.} The following statement in \cite{Grebenev17} on p.$\,$8, that {\it ``the relationships (46) also demonstrate the exceptional role of the zero-vorticity constraint (32) to guarantee that $c^{11}$ and $c^{12}$ are harmonic functions"}, is misleading. In Sec.$\,$\ref{Sec.2.3} we clearly demonstrated that the choice~$\omega=0$ is not exceptional at all, because the obtained invariances can always be equivalently re-formulated such that {\it any} arbitrary but fixed value of $\omega\in\mathbb{R}$ will do the same job as the particular choice $\omega=0$ --- see again the result \eqref{180213:1725} for an alternative $\xi^3$ and its subsequent discussion. Hence, the choice of a zero-vorticity constraint $\omega=0$ plays no exceptional role, with the result again of Sec.$\,$\ref{Sec.2.3} that the proposed conformal invariance is not only broken for $\omega=0$, but for all $\omega\in\mathbb{R}$, thus refuting \cite{Grebenev17} in its most general form.

\noindent {\bf R4.} Important to note in this overall discussion is that the invariant transformation examined in this investigation \eqref{180201:1057}-\eqref{180201:1703}\footnote{Note that only the reduced case $c^{11}_1=c^{11}_2=0$ constitutes an invariant transformation.} is only an equivalence and not a true symmetry transformation, simply due to that we are dealing here with an unclosed system of equations \eqref{180130:2115}-\eqref{180130:2158} where the dynamical rule for the 2-point PDF $f_2$ is not known beforehand.
In contrast to a true\linebreak[4] symmetry transformation, which maps a solution of a specific (closed) equation to a new solution of the same equation, an equivalence transform acts in a weaker sense in that it only maps an (unclosed) equation to a new (unclosed) equation of the same class.\footnote{Equivalence transformations can be successfully applied for example to classify unclosed differential equations according to the number of symmetries they admit when specifying the unclosed terms (see e.g. \cite{Meleshko02,Khabirov02.1,Khabirov02.2,Chirkunov12,Meleshko15,Bihlo17}). A typical task in this context sometimes is to find a specification of the unclosed terms such that the maximal symmetry algebra is gained. Once the equation is closed by a such a group classification, invariant solutions can be determined. But in how far these equations and their solutions are physically relevant and whether they can be matched to empirical data is not clarified {\it a priori} by this approach, in particular if such a pure Lie-group-based type of modelling is performed fully detached from empirical research. In this regard, special attention has to be given to the unclosed statistical equations of turbulence as considered herein, since the unclosed 2-point PDF in \eqref{180130:2115}-\eqref{180130:2158}\linebreak[4] is only an analytical and theoretical unknown, but not an empirical one since it is fully determined by the underlying deterministic Navier-Stokes equations, which again are well-known for to break statistical symmetries in turbulence within intermittent events (see e.g. \cite{Frisch95}). Hence extra caution has to be exercised when employing a pure symmetry-based modeling to turbulence.} Of course, it is trivial and

\restoregeometry

\newgeometry{left=2.50cm,right=2.44cm,top=2.50cm,bottom=2.50cm,headsep=1em}

\noindent goes without saying that if once a real solution for $f_2$ is known, and if the equivalence \eqref{180203:1632} itself is physically realizable,\footnote{The equivalence transformation
\eqref{180203:1632} is said to be physically realizable if the transformed field $\tilde{f}_2$ can be generated as a PDF-solution of the deterministic Navier-Stokes equations according to its transformation
rule $\tilde{f}_2=f_2+\epsilon\cdot\eta_{f_2}+\mathcal{O}(\epsilon^2)$, where we assume that the non-transformed field $f_2$ already constitutes a PDF-solution.
In other words, if the transformed $\tilde{f}_2$ cannot emerge dynamically from the non-transformed $f_2$ via the deterministic and thus closed Navier-Stokes equations, then the equivalence \eqref{180203:1632} is nonphysical.
To prove whether \eqref{180203:1632} is physically realizable or not, is beyond the scope of this article. However, there are a few examples of statistical Navier-Stokes equivalences which are clearly nonphysical
--- see e.g. \cite{Frewer14.2,Frewer15.1,Frewer16.1,Frewer17,Sadeghi20}.}
then this equivalence turns into a symmetry transformation and $f_2$ gets mapped to a new solution $\tilde{f}_2=f_2+\epsilon\cdot\eta_{f_2}+\mathcal{O}(\epsilon^2)$. But since this is not the case here, any invariant transformation of \eqref{180201:1057}-\eqref{180201:1703} will thus at this stage only map between equations and not between solutions, where $f_2$ then is the unknown source or sink term, or collectively the unknown constitutive law of these equations.

Hence, for the global invariant scaling determined herein ($c^{11}_1=c^{11}_2=0$) we cannot expect any information about the inner solution structure of the 1-point PDF equation as long as the dynamical equation for the 2-point PDF $f_2$ is not modeled.
Without empirical modeling it is clear that the closure problem of turbulence cannot be circumvented by just employing the method of a Lie-group symmetry analysis. For more details on this issue, see e.g. \cite{Frewer14.1,Frewer14.2} and the references therein.

\noindent {\bf R5.}$\;\,$Finally, it should not go unmentioned that \cite{Grebenev17}$\,$\&$\,$\cite{Waclawczyk17}\linebreak[4] are not the first articles from the group of Oberlack {\it et al.} dealing with symmetries and the LMN equations which are flawed. The previously published comments by \cite{Frewer14.2,Frewer15.1,Frewer15.2,Frewer16.1,Frewer17} and \cite{Frewer16.2} clearly prove this. Nor should it be ignored that the present
flawed~result of \cite{Grebenev17} forms a basic building block of a recently granted 3-year DFG project (Gepris, No.$\,$385665358 \href{http://gepris.dfg.de/gepris/projekt/385665358?language=en}{\ExternalLink}$\,$). A detailed critical discussion of this project is given in ResearchGate \href{https://www.researchgate.net/project/Turbulence-and-Symmetries/update/5a33bec64cde266d587b3b78}{\ExternalLink}$\,$. In this regard please also visit \url{https://zenodo.org/communities/turbsym/}.

\appendix

\section{The general and full invariance group of the local equation $\boldsymbol{\mathsf{E}}_{\boldsymbol{1}}$~(1.4)\label{Sec.A}}

We present two different but equivalent versions as how one can perform a systematic and complete Lie-group invariance analysis of the local equation \eqref{180131:0807} using a software package,\linebreak[4]
the DESOLV-II package of \cite{Vu12}.

In the first version (Version~No.$\,$1), we consider the three dependent variables $J^0$, $J^1$, $J^2$ in \eqref{180131:0807} to explicitly depend on all independent variables of the system involved. These are seven in total and are listed in \eqref{180201:1054}. Although we only consider here the local equation \eqref{180131:0807} and ignore in this step the non-local equations \eqref{180201:1004}-\eqref{180201:1007}, we nevertheless should provide the symmetry searching algorithm with the information that three independent variables, namely $y^4=x^\prime$, $y^5=y^\prime$, $y^6=\omega^\prime$, are integration variables. This can be done by augmenting the local equation~\eqref{180131:0807} with first-order differential consequences consistent with all equations \eqref{180131:0807}-\eqref{180201:1007} defining the system. The relevant ones are given by the system of equations $\partial_{y^j}J^k=0$, for all $k=0,1,2$ and $j=4,5,6$, obviously telling us that all three dependent variables $J^0$, $J^1$, $J^2$ do not explicitly dependent on the three (integration) variables $y^4$, $y^5$, $y^6$, which brings us to the second~version.

In the second version (Version~No.$\,$2), we consider the three dependent variables $J^0$, $J^1$, $J^2$ in~\eqref{180131:0807} to explicitly depend only on those independent variables which the full system \eqref{180131:0807}-\eqref{180201:1007} defines for them. As established in the first version above, they thus can only dependent on the four variables $y^0$, $y^1$, $y^2$, $y^3$. Hence, the symmetry analysis of this version will only involve a single equation, the local equation \eqref{180131:0807} itself.

As the computer results show below, {\it both} versions obviously yield exactly the same final result \eqref{180201:0924}-\eqref{180201:0930}, in that the infinitesimal $\xi^3$ is only a function of $\omega$, and this without any restrictions on the values of $\omega$, as also stated correctly by~\eqref{190712:1711}.

\restoregeometry

\subsection{ Version No.$\,$1}
\input{V1}

\subsection{ Version No.$\,$2}
\input{V2}
\bibliographystyle{jfm}
\bibliography{References}

\end{document}

%% file: V1.tex
\begin{maplegroup}
\begin{flushleft}
{\large Header:}
\end{flushleft}
\end{maplegroup}
\begin{maplegroup}
\begin{mapleinput}
\mapleinline{active}{1d}{restart: read "Desolv-V5R5.mpl": with(desolv):}{}
\end{mapleinput}
\mapleresult
\begin{maplelatex}
\mapleinline{inert}{2d}{`DESOLVII_V5R5 (March-2011)(c) by Dr. K.
T. Vu, Dr. J. Carminati and
Miss. G. Jefferson`;}{%
\maplemultiline{ \mathit{\phantom{xxxxxxxx} DESOLVII\_V5R5\ (March-2011)(c)} \\
\mathit{by\ Dr.\ K.\ T.\ Vu,\ Dr.\ J.\ Carminati\ and\ Miss.\ G.\
Jefferson} }}
\end{maplelatex}
\end{maplegroup}
\begin{maplegroup}
\begin{flushleft}
{\large Definitions of variables, local equation and differential consequences:}
\end{flushleft}
\end{maplegroup}
\begin{maplegroup}
\begin{mapleinput}
\mapleinline{active}{1d}{alias(sigma=(y0,y1,y2,y3,y4,y5,y6,J0,J1,J2)): Y:=(y0,y1,y2,y3,y4,y5,y6):
}{}
\end{mapleinput}
\end{maplegroup}
\begin{maplegroup}
\begin{mapleinput}
\mapleinline{active}{1d}{eqn0:=diff(J0(Y),y0)+diff(J1(Y),y1)+diff(J2(Y),y2)=0:
}{}
\end{mapleinput}
\end{maplegroup}
\begin{maplegroup}
\begin{mapleinput}
\mapleinline{active}{1d}{eqn1:=diff(J0(Y),y4)=0: eqn2:=diff(J0(Y),y5)=0: eqn3:=diff(J0(Y),y6)=0:
\hspace{0.43cm} eqn4:=diff(J1(Y),y4)=0: eqn5:=diff(J1(Y),y5)=0: eqn6:=diff(J1(Y),y6)=0:
\hspace{0.43cm} eqn7:=diff(J2(Y),y4)=0: eqn8:=diff(J2(Y),y5)=0: eqn9:=diff(J2(Y),y6)=0:
}{}
\end{mapleinput}
\end{maplegroup}
\begin{maplegroup}
\begin{mapleinput}
\mapleinline{active}{1d}{eqns:=[eqn0,eqn1,eqn2,eqn3,eqn4,eqn5,eqn6,eqn7,eqn8,eqn9]:
}{}
\end{mapleinput}
\end{maplegroup}
\begin{maplegroup}
\begin{flushleft}
{\large Symmetry Algorithm:}\\[0.75em]
\textit{{\large Size of the determining system:}}
\end{flushleft}
\end{maplegroup}
\begin{maplegroup}
\begin{mapleinput}
\mapleinline{active}{1d}{detsys:=gendef(eqns,[J0,J1,J2],[y0,y1,y2,y3,y4,y5,y6]): nops(detsys[1]);
}{}
\end{mapleinput}
\mapleresult
\begin{maplelatex}
\mapleinline{inert}{2d}{45}{\[\displaystyle 45\]}
\end{maplelatex}
\end{maplegroup}
\vspace{-0.75em}
\begin{maplegroup}
\begin{flushleft}
\textit{{\large Solving the determining system:}}
\end{flushleft}
\end{maplegroup}
\begin{maplegroup}
\begin{mapleinput}
\mapleinline{active}{1d}{sym:=pdesolv(op(detsys));
}{}
\end{mapleinput}
\mapleresult
\begin{maplelatex}
\mapleinline{inert}{2d}{}{\[\displaystyle
{\it sym}\, := \,\bigg[\bigg[-{\frac {\partial }{\partial {\it y1}}}{\it F\_15} \left( {\it y0},{\it y1},{\it y2},{\it y3}\\ \mbox{} \right)
-{\frac {\partial }{\partial {\it y2}}}{\it F\_21} \left( {\it y0},{\it y1},{\it y2},{\it y3}\\
\mbox{} \right) -{\it F\_47} \left( {\it y0},{\it y1},{\it y2},{\it y3}\\ \mbox{} \right),
\]}
\end{maplelatex}
\mapleresult
\begin{maplelatex}
\mapleinline{inert}{2d}{}{\[\displaystyle
{\frac {\partial }{\partial {\it y0}}}{\it F\_44} \left( {\it y0},{\it y1},{\it y2},{\it y3}\\
\mbox{} \right)
+{\frac {\partial }{\partial {\it y1}}}{\it F\_45} \left( {\it y0},{\it y1},{\it y2},{\it y3}\\
\mbox{} \right)
+{\frac {\partial }{\partial {\it y2}}}{\it F\_46} \left( {\it y0},{\it y1},{\it y2},{\it y3}\\
\mbox{} \right) \\
\mbox{} \bigg],\,[\,],
\]}
\end{maplelatex}
\mapleresult
\begin{maplelatex}
\mapleinline{inert}{2d}{}{\[\displaystyle
\bigg[\xi_{{{\it y0}}} \left( {\it \sigma}\right)\!=\!{\it F\_27}\\
\mbox{} \left( {\it y0},{\it y1},{\it y2},{\it y3} \right),
\xi_{{{\it y1}}} \left( {\it \sigma}\right)\!=\!{\it F\_15} \left( {\it y0},{\it y1},{\it y2},{\it y3} \right),
\xi_{{{\it y2}}} \left( {\it \sigma}\right)\\
\mbox{}\!=\!{\it F\_21} \left( {\it y0},{\it y1},{\it y2},{\it y3} \right),
\]}
\end{maplelatex}
\mapleresult
\begin{maplelatex}
\mapleinline{inert}{2d}{}{\[\displaystyle
\xi_{{{\it y3}}} \left( {\it \sigma}\right)\!=\!{\it F\_9} \left( {\it y3} \right),
\xi_{{{\it y4}\\
\mbox{}}} \left( {\it \sigma}\right) \\
\mbox{}\!=\!\xi_{{{\it y4}\\
\mbox{}}} \left( {\it \sigma}\right),
\xi_{{{\it y5}}} \left( {\it \sigma}\right)\!=\!\xi_{{{\it y5}}} \left( {\it \sigma}\right),
\xi_{{{\it y6}}} \left( {\it \sigma}\right)\!=\!\xi_{{{\it y6}}} \left( {\it \sigma}\right),
\]}
\end{maplelatex}
\mapleresult
\begin{maplelatex}
\mapleinline{inert}{2d}{}{\[\displaystyle
\eta_{{{\it J0}}} \left( {\it \sigma}\right)\!=\!
{\it F\_47} \left( {\it y0},{\it y1},{\it y2},{\it y3} \right){\it J0}
\!+\!{\it J1}\,{\frac {\partial }{\partial {\it y1}}}{\it F\_27}\\
\mbox{} \left( {\it y0},{\it y1},{\it y2},{\it y3} \right)
\]}
\end{maplelatex}
\mapleresult
\vspace{-0.5em}
\begin{maplelatex}
\mapleinline{inert}{2d}{}{\[\displaystyle
\hspace{1.5cm}+{\it J2}\,{\frac {\partial }{\partial {\it y2}}}{\it F\_27}\\
\mbox{} \left( {\it y0},{\it y1},{\it y2},{\it y3} \right)\!+\!{\it F\_44} \left( {\it y0},{\it y1},{\it y2},{\it y3} \right) \\
\mbox{},
\]}
\end{maplelatex}
\vspace{-0.5em}
\mapleresult
\begin{maplelatex}
\mapleinline{inert}{2d}{}{\[\displaystyle
\eta_{{{\it J1}}} \left( {\it \sigma}\right)\!=\!
{\it J0}\,{\frac {\partial }{\partial {\it y0}}}{\it F\_15} \left( {\it y0},{\it y1},{\it y2},{\it y3} \right) \\
\mbox{}\!+\!{\it F\_47} \left( {\it y0},{\it y1},{\it y2},{\it y3} \right) {\it J1}
\!+\!{\it J2}\,{\frac {\partial }{\partial {\it y2}}}{\it F\_15} \left( {\it y0},{\it y1},{\it y2},{\it y3} \right)
\]}
\end{maplelatex}
\mapleresult
\vspace{-0.25em}
\begin{maplelatex}
\mapleinline{inert}{2d}{}{\[\displaystyle
\hspace{1.5cm}+{\it J1}\,{\frac {\partial }{\partial {\it y1}}}{\it F\_15} \left( {\it y0},{\it y1},{\it y2},{\it y3} \right)
\!-\!{\it J1}\,{\frac {\partial }{\partial {\it y0}}}{\it F\_27}\\
\mbox{} \left( {\it y0},{\it y1},{\it y2},{\it y3} \right)\!+\!{\it F\_45} \left( {\it y0},{\it y1},{\it y2},{\it y3} \right) ,
\]}
\end{maplelatex}
\mapleresult
\begin{maplelatex}
\mapleinline{inert}{2d}{}{\[\displaystyle
\eta_{{{\it J2}}} \left( {\it \sigma}\right) \\
\mbox{}\!=\!
{\it J0}\,{\frac {\partial }{\partial {\it y0}}}{\it F\_21} \left( {\it y0},{\it y1},{\it y2},{\it y3} \right)
\!+\!\,{\it F\_47} \left( {\it y0},{\it y1},{\it y2},{\it y3} \right){\it J2}
\!+\!{\it J2}\,{\frac {\partial }{\partial {\it y2}}}{\it F\_21} \left( {\it y0},{\it y1},{\it y2},{\it y3} \right)
\]}
\end{maplelatex}
\mapleresult
\begin{maplelatex}
\mapleinline{inert}{2d}{}{\[\displaystyle
\hspace{1.5cm}+{\it J1}\,{\frac {\partial }{\partial {\it y1}}}{\it F\_21} \left( {\it y0},{\it y1},{\it y2},{\it y3} \right)
\!-\!{\it J2}\,{\frac {\partial }{\partial {\it y0}}}{\it F\_27}\\
\mbox{} \left( {\it y0},{\it y1},{\it y2},{\it y3} \right)
\!+\!{\it F\_46} \left( {\it y0},{\it y1},{\it y2},{\it y3} \right)\\
\mbox{}\bigg],
\]}
\end{maplelatex}
\mapleresult
\begin{maplelatex}
\mapleinline{inert}{2d}{}{\[\displaystyle
\bigg[{\it F\_15} \left( {\it y0},{\it y1},{\it y2},{\it y3}\\
\mbox{} \right) ,{\it F\_21} \left( {\it y0},{\it y1},{\it y2},{\it y3}\\
\mbox{} \right) ,{\it F\_27} \left( {\it y0},{\it y1},{\it y2},{\it y3}\\
\mbox{} \right) ,{\it F\_44} \left( {\it y0},{\it y1},{\it y2},{\it y3}\\
\mbox{} \right), \\
\mbox{}
\]}
\end{maplelatex}
\mapleresult
\begin{maplelatex}
\mapleinline{inert}{2d}{}{\[\displaystyle
{\it F\_45} \left( {\it y0},{\it y1},{\it y2},{\it y3}\\
\mbox{} \right) ,{\it F\_46} \left( {\it y0},{\it y1},{\it y2},{\it y3}\\
\mbox{} \right) ,{\it F\_47} \left( {\it y0},{\it y1},{\it y2},{\it y3}\\
\mbox{} \right) \\
\mbox{},
{\it F\_9} \left( {\it y3}\\
\mbox{} \right),
\]}
\end{maplelatex}
\mapleresult
\begin{maplelatex}
\mapleinline{inert}{2d}{}{\[\displaystyle
\xi_{{{\it y4}}} \left( {\it \sigma}\right),
\xi_{{{\it y5}}} \left( {\it \sigma}\right),
\xi_{{{\it y6}\\
\mbox{}}} \left( {\it \sigma}\right) \\
\mbox{}\bigg]\bigg]
\]}
\end{maplelatex}
\end{maplegroup}
\begin{maplegroup}
\begin{flushleft}
{\large Redefining solution functions as used in (1.9)-(1.11):}
\end{flushleft}
\end{maplegroup}
\begin{maplegroup}
\begin{mapleinput}
\mapleinline{active}{1d}{F_27(y0,y1,y2,y3):=xi0(y0,y1,y2,y3); F_15(y0,y1,y2,y3):=xi1(y0,y1,y2,y3);
\hspace{0.43cm} F_21(y0,y1,y2,y3):=xi2(y0,y1,y2,y3); F_9(y3):=xi3(y3);
\hspace{0.43cm} F_47(y0,y1,y2,y3):=-diff(F_15(y0,y1,y2,y3),y1)-diff(F_21(y0,y1,y2,y3),y2);
\hspace{0.43cm} F_44(y0,y1,y2,y3):=b0(y0,y1,y2,y3)-C(y3)*j0(y0,y1,y2,y3);
\hspace{0.43cm} F_45(y0,y1,y2,y3):=b1(y0,y1,y2,y3)-C(y3)*j1(y0,y1,y2,y3);
\hspace{0.43cm} F_46(y0,y1,y2,y3):=b2(y0,y1,y2,y3)-C(y3)*j2(y0,y1,y2,y3);
}{}
\end{mapleinput}
\mapleresult
\begin{maplelatex}
\mapleinline{inert}{2d}{}{\[\displaystyle
{\it F\_27} \left({\it y0} ,{\it y1} ,{\it y2} ,{\it y3} \right):={\it \xi 0}\\
\mbox{} \left({\it y0} ,{\it y1} ,{\it y2} ,{\it y3} \right)
\]}
\end{maplelatex}
\mapleresult
\begin{maplelatex}
\mapleinline{inert}{2d}{}{\[\displaystyle
{\it F\_15} \left({\it y0} ,{\it y1} ,{\it y2} ,{\it y3} \right):={\it \xi 1}\\
\mbox{} \left({\it y0} ,{\it y1} ,{\it y2} ,{\it y3} \right)
\]}
\end{maplelatex}
\mapleresult
\begin{maplelatex}
\mapleinline{inert}{2d}{}{\[\displaystyle
{\it F\_21} \left({\it y0} ,{\it y1} ,{\it y2} ,{\it y3} \right):={\it \xi 2}\\
\mbox{} \left({\it y0} ,{\it y1} ,{\it y2} ,{\it y3} \right)
\]}
\end{maplelatex}
\mapleresult
\begin{maplelatex}
\mapleinline{inert}{2d}{}{\[\displaystyle
\hspace{0.20cm}{\it F\_9} \left({\it y3} \right):={\it \xi 3}\\
\mbox{} \left({\it y3} \right)
\]}
\end{maplelatex}
\mapleresult
\begin{maplelatex}
\mapleinline{inert}{2d}{}{\[\displaystyle
{\it F\_47} \left({\it y0} ,{\it y1} ,{\it y2} ,{\it y3} \right):=-\frac{\partial }{\partial {\it y1} }~{\it \xi 1}\\
\mbox{} \left({\it y0} ,{\it y1} ,{\it y2} ,{\it y3} \right)-\frac{\partial }{\partial {\it y2} }~{\it \xi 2}\\
\mbox{} \left({\it y0} ,{\it y1} ,{\it y2} ,{\it y3} \right)
\]}
\end{maplelatex}
\mapleresult
\begin{maplelatex}
\mapleinline{inert}{2d}{}{\[\displaystyle
{\it F\_44} \left({\it y0} ,{\it y1} ,{\it y2} ,{\it y3} \right):={\it b0}\\
\mbox{} \left({\it y0} ,{\it y1} ,{\it y2} ,{\it y3} \right)-C \left({\it y3} \right)~{\it j0} \left({\it y0} ,{\it y1} ,{\it y2} ,{\it y3} \right)
\]}
\end{maplelatex}
\mapleresult
\begin{maplelatex}
\mapleinline{inert}{2d}{}{\[\displaystyle
{\it F\_45} \left({\it y0} ,{\it y1} ,{\it y2} ,{\it y3} \right):={\it b1}\\
\mbox{} \left({\it y0} ,{\it y1} ,{\it y2} ,{\it y3} \right)-C \left({\it y3} \right)~{\it j1} \left({\it y0} ,{\it y1} ,{\it y2} ,{\it y3} \right)
\]}
\end{maplelatex}
\mapleresult
\begin{maplelatex}
\mapleinline{inert}{2d}{}{\[\displaystyle
{\it F\_46} \left({\it y0} ,{\it y1} ,{\it y2} ,{\it y3} \right):={\it b2}\\
\mbox{} \left({\it y0} ,{\it y1} ,{\it y2} ,{\it y3} \right)-C \left({\it y3} \right)~{\it j2} \left({\it y0} ,{\it y1} ,{\it y2} ,{\it y3} \right)
\]}
\end{maplelatex}
\end{maplegroup}
\begin{maplegroup}
\begin{flushleft}
{\large The arbitrary integration functions $b_k$ and $j_k$ are solutions of the local equation (1.4).\\
Since (1.4) is a linear equation: if $b_k$ and $j_k$ are solutions, so is $B_k:=b_k-C\cdot j_k$.}
\end{flushleft}
\end{maplegroup}
\begin{maplegroup}
\begin{mapleinput}
\mapleinline{active}{1d}{simplify(sym[1,1]); simplify(sym[1,2])=0;}{}
\end{mapleinput}
\mapleresult
\begin{maplelatex}
\mapleinline{inert}{2d}{0}{\[\displaystyle 0\]}
\end{maplelatex}
\mapleresult
\begin{maplelatex}
\mapleinline{inert}{2d}{}{\[\displaystyle
{\frac {\partial }{\partial {\it y0}}}{\it b0} \left( {\it y0},{\it y1},{\it y2},{\it y3} \right)
+{\frac {\partial }{\partial {\it y1}}}{\it b1} \left( {\it y0},{\it y1},{\it y2},{\it y3} \right)
+{\frac {\partial }{\partial {\it y2}}}{\it b2} \left( {\it y0},{\it y1},{\it y2},{\it y3} \right)
\]}
\end{maplelatex}
\mapleresult
\begin{maplelatex}
\mapleinline{inert}{2d}{}{\[\displaystyle
-C \left( {\it y3} \right)\left(
{\frac {\partial }{\partial {\it y0}}}{\it j0} \left( {\it y0},{\it y1},{\it y2},{\it y3} \right)
+{\frac {\partial }{\partial {\it y1}}}{\it j1} \left( {\it y0},{\it y1},{\it y2},{\it y3} \right)
+{\frac {\partial }{\partial {\it y2}}}{\it j2} \left( {\it y0},{\it y1},{\it y2},{\it y3} \right)
\right)=0
\]}
\end{maplelatex}
\end{maplegroup}
%
%
\begin{maplegroup}
\begin{flushleft}
{\large The solutions $j^k$ can be identified as the dependent variables $J^k$:}
\end{flushleft}
\end{maplegroup}
\begin{maplegroup}
\begin{mapleinput}
\mapleinline{active}{1d}{j0(y0,y1,y2,y3):=J0: j1(y0,y1,y2,y3):=J1: j2(y0,y1,y2,y3):=J2:
}{}
\end{mapleinput}
\end{maplegroup}
\begin{maplegroup}
\begin{flushleft}
{\large Final result which is identical to (1.9)-(1.11):}
\end{flushleft}
\end{maplegroup}
\begin{maplegroup}
\begin{mapleinput}
\mapleinline{active}{1d}{simplify(sym[3,1]); simplify(sym[3,2]); simplify(sym[3,3]);
\hspace{0.43cm} simplify(sym[3,4]); simplify(sym[3,8]); simplify(sym[3,9]);
\hspace{0.43cm} simplify(sym[3,10]);
}{}
\end{mapleinput}
\mapleresult
\begin{maplelatex}
\mapleinline{inert}{2d}{}{\[\displaystyle \xi_{{{\it y0}}} \left( {\it \sigma}\right)
={\it \xi 0}\\
\mbox{} \left( {\it y0},{\it y1},{\it y2},{\it y3} \right) \]}
\end{maplelatex}
\mapleresult
\begin{maplelatex}
\mapleinline{inert}{2d}{}{\[\displaystyle \xi_{{{\it y1}}} \left( {\it \sigma}\right) ={\it \xi 1}\\
\mbox{} \left( {\it y0},{\it y1},{\it y2},{\it y3} \right) \]}
\end{maplelatex}
\mapleresult
\begin{maplelatex}
\mapleinline{inert}{2d}{}{\[\displaystyle \xi_{{{\it y2}}} \left( {\it \sigma}\right) ={\it \xi 2}\\
\mbox{} \left( {\it y0},{\it y1},{\it y2},{\it y3} \right) \]}
\end{maplelatex}
\mapleresult
\begin{maplelatex}
\mapleinline{inert}{2d}{}{\[\displaystyle \xi_{{{\it y3}}} \left( {\it \sigma}\right) ={\it \xi 3}\\
\mbox{} \left( {\it y3} \right) \]}
\end{maplelatex}
\mapleresult
\begin{maplelatex}
\mapleinline{inert}{2d}{}{\[\displaystyle
\eta_{{{\it J0}}} \left( {\it \sigma}\right)=
{\it J1}\,{\frac {\partial }{\partial {\it y1}}}{\it \xi 0} \left( {\it y0},{\it y1},{\it y2},{\it y3}\\\mbox{} \right)
+{\it J2}\,{\frac {\partial }{\partial {\it y2}}}{\it \xi 0} \left( {\it y0},{\it y1},{\it y2},{\it y3}\\\mbox{} \right) \\\mbox{}
-{\it J0}\,{\frac {\partial }{\partial {\it y1}}}{\it \xi 1} \left( {\it y0},{\it y1},{\it y2},{\it y3}\\\mbox{} \right) \\\mbox{}
\]}
\end{maplelatex}
\mapleresult
\begin{maplelatex}
\mapleinline{inert}{2d}{}{\[\displaystyle
-{\it J0}\,{\frac {\partial }{\partial {\it y2}}}{\it \xi 2} \left( {\it y0},{\it y1},{\it y2},{\it y3}\\\mbox{} \right)
-C \left( {\it y3}\\\mbox{} \right) {\it J0}
+{\it b0} \left( {\it y0},{\it y1},{\it y2},{\it y3}\\\mbox{} \right)
\]}
\end{maplelatex}
\mapleresult
\begin{maplelatex}
\mapleinline{inert}{2d}{}{\[\displaystyle
\eta_{{{\it J1}}} \left( {\it \sigma}\right)=
{\it J0}\,{\frac {\partial }{\partial {\it y0}}}{\it \xi 1} \left( {\it y0},{\it y1},{\it y2},{\it y3}\\\mbox{} \right)
+{\it J2}\,{\frac {\partial }{\partial {\it y2}}}{\it \xi 1} \left( {\it y0},{\it y1},{\it y2},{\it y3}\\\mbox{} \right) \\\mbox{}
-{\it J1}\,{\frac {\partial }{\partial {\it y0}}}{\it \xi 0} \left( {\it y0},{\it y1},{\it y2},{\it y3}\\\mbox{} \right)
\]}
\end{maplelatex}
\mapleresult
\begin{maplelatex}
\mapleinline{inert}{2d}{}{\[\displaystyle
-{\it J1}\,{\frac {\partial }{\partial {\it y2}}}{\it \xi 2} \left( {\it y0},{\it y1},{\it y2},{\it y3}\\\mbox{} \right)
-C \left( {\it y3}\\\mbox{} \right) {\it J1}
+{\it b1} \left( {\it y0},{\it y1},{\it y2},{\it y3}\\\mbox{} \right) \\\mbox{}
\]}
\end{maplelatex}
\mapleresult
\begin{maplelatex}
\mapleinline{inert}{2d}{}{\[\displaystyle
\eta_{{{\it J2}}} \left( {\it \sigma}\right)=
{\it J0}\,{\frac {\partial }{\partial {\it y0}}}{\it \xi 2} \left( {\it y0},{\it y1},{\it y2},{\it y3}\\\mbox{} \right)
+{\it J1}\,{\frac {\partial }{\partial {\it y1}}}{\it \xi 2} \left( {\it y0},{\it y1},{\it y2},{\it y3}\\\mbox{} \right) \\\mbox{}
-{\it J2}\,{\frac {\partial }{\partial {\it y0}}}{\it \xi 0} \left( {\it y0},{\it y1},{\it y2},{\it y3}\\\mbox{} \right)
\]}
\end{maplelatex}
\mapleresult
\begin{maplelatex}
\mapleinline{inert}{2d}{}{\[\displaystyle
-{\it J2}\,{\frac {\partial }{\partial {\it y1}}}{\it \xi 1} \left( {\it y0},{\it y1},{\it y2},{\it y3}\\\mbox{} \right)
-C \left( {\it y3}\\\mbox{} \right) {\it J2}
+{\it b2} \left( {\it y0},{\it y1},{\it y2},{\it y3}\\\mbox{} \right) \\\mbox{}
\]}
\end{maplelatex}
\end{maplegroup}
%
%
\begin{maplegroup}
\begin{flushleft}
{\large Counter-checking above result if truly a solution --- and indeed it is:}
\end{flushleft}
\end{maplegroup}
\begin{maplegroup}
\begin{mapleinput}
\mapleinline{active}{1d}{rho:=(y0,y1,y2,y3):}{}
\end{mapleinput}
\begin{mapleinput}
\mapleinline{active}{1d}{xi[y0](sigma):=xi0(rho); xi[y1](sigma):=xi1(rho);
\hspace{0.43cm} xi[y2](sigma):=xi2(rho); xi[y3](sigma):=xi3(y3);
\hspace{0.43cm} eta[J0](sigma):=J1*(diff(xi0(rho),y1))+J2*(diff(xi0(rho),y2))
\hspace{0.43cm} -J0*(diff(xi1(rho),y1))-J0*(diff(xi2(rho),y2))-C(y3)*J0+b0(rho);
\hspace{0.43cm} eta[J1](sigma):=J0*(diff(xi1(rho),y0))+J2*(diff(xi1(rho),y2))
\hspace{0.43cm} -J1*(diff(xi0(rho),y0))-J1*(diff(xi2(rho),y2))-C(y3)*J1+b1(rho);
\hspace{0.43cm} eta[J2](sigma):=J0*(diff(xi2(rho),y0))+J1*(diff(xi2(rho),y1))
\hspace{0.43cm} -J2*(diff(xi0(rho),y0))-J2*(diff(xi1(rho),y1))-C(y3)*J2+b2(rho);
}{}
\end{mapleinput}
\mapleresult
\begin{maplelatex}
\mapleinline{inert}{2d}{}{\[\displaystyle
\xi_{{{\it y0}}} \left( {\it \sigma}\right)={\it \xi 0}\\\mbox{} \left(\rho \right),\;
\xi_{{{\it y1}}} \left( {\it \sigma}\right) ={\it \xi 1}\\\mbox{} \left( \rho \right),\;
\xi_{{{\it y2}}} \left( {\it \sigma}\right) ={\it \xi 2}\\\mbox{} \left( \rho \right),\;
\xi_{{{\it y3}}} \left( {\it \sigma}\right) ={\it \xi 3}\\\mbox{} \left( {\it y3} \right)
\]}
\end{maplelatex}
\mapleresult
\begin{maplelatex}
\mapleinline{inert}{2d}{}{\[\displaystyle
\eta_{{{\it J0}}} \left( {\it \sigma}\right)=
{\it J1}\,{\frac {\partial }{\partial {\it y1}}}{\it \xi 0} \left( \rho\\\mbox{} \right)
\!+\!{\it J2}\,{\frac {\partial }{\partial {\it y2}}}{\it \xi 0} \left(\rho\\\mbox{} \right) \\\mbox{}
\!-\!{\it J0}\,{\frac {\partial }{\partial {\it y1}}}{\it \xi 1} \left( \rho\\\mbox{} \right) \\\mbox{}
\!-\!{\it J0}\,{\frac {\partial }{\partial {\it y2}}}{\it \xi 2} \left( \rho\\\mbox{} \right)
\!-\!C \left( {\it y3}\\\mbox{} \right) {\it J0}
\!+\!{\it b0} \left(\rho\\\mbox{} \right)
\]}
\end{maplelatex}
\mapleresult
\begin{maplelatex}
\mapleinline{inert}{2d}{}{\[\displaystyle
\eta_{{{\it J1}}} \left( {\it \sigma}\right)=
{\it J0}\,{\frac {\partial }{\partial {\it y0}}}{\it \xi 1} \left( \rho\\\mbox{} \right)
\!+\!{\it J2}\,{\frac {\partial }{\partial {\it y2}}}{\it \xi 1} \left( \rho\\\mbox{} \right) \\\mbox{}
\!-\!{\it J1}\,{\frac {\partial }{\partial {\it y0}}}{\it \xi 0} \left( \rho\\\mbox{} \right)
\!-\!{\it J1}\,{\frac {\partial }{\partial {\it y2}}}{\it \xi 2} \left( \rho\\\mbox{} \right)
\!-\!C \left( {\it y3}\\\mbox{} \right) {\it J1}
\!+\!{\it b1} \left( \rho\\\mbox{} \right) \\\mbox{}
\]}
\end{maplelatex}
\mapleresult
\begin{maplelatex}
\mapleinline{inert}{2d}{}{\[\displaystyle
\eta_{{{\it J2}}} \left( {\it \sigma}\right)=
{\it J0}\,{\frac {\partial }{\partial {\it y0}}}{\it \xi 2} \left( \rho\\\mbox{} \right)
\!+\!{\it J1}\,{\frac {\partial }{\partial {\it y1}}}{\it \xi 2} \left( \rho\\\mbox{} \right) \\\mbox{}
\!-\!{\it J2}\,{\frac {\partial }{\partial {\it y0}}}{\it \xi 0} \left( \rho\\\mbox{} \right)
\!-\!{\it J2}\,{\frac {\partial }{\partial {\it y1}}}{\it \xi 1} \left( \rho\\\mbox{} \right)
\!-\!C \left( {\it y3}\\\mbox{} \right) {\it J2}
\!+\!{\it b2} \left( \rho\\\mbox{} \right) \\\mbox{}
\]}
\end{maplelatex}
\vspace{1em}
\begin{mapleinput}
\mapleinline{active}{1d}{simplify(detsys[1]);
}{}
\end{mapleinput}
\mapleresult
\begin{maplelatex}
\mapleinline{inert}{2d}{}{\[\displaystyle
\bigg[0,0,0,0,0,0,0,0,0,0,0,0,0,0,0,0,0,0,0,0,0,0,0,0,0,0,0,0,0,0,0,
\]}
\end{maplelatex}
\mapleresult
\begin{maplelatex}
\mapleinline{inert}{2d}{}{\[\displaystyle
\hspace{0.1cm} 0,0,0,0,0,0,0,0,0,0,0,
{\frac {\partial }{\partial {\it y0}}}{\it b0} \left( \rho \right)
+{\frac {\partial }{\partial {\it y1}}}{\it b1} \left( \rho \right)
+{\frac {\partial }{\partial {\it y2}}}{\it b2} \left( \rho\right) \\\mbox{} ,0,0 \bigg]
\]}
\end{maplelatex}
\end{maplegroup}

%% file: V2.tex
\begin{maplegroup}
\begin{flushleft}
{\large Header:}
\end{flushleft}
\end{maplegroup}
\begin{maplegroup}
\begin{mapleinput}
\mapleinline{active}{1d}{restart: read "Desolv-V5R5.mpl": with(desolv):}{}
\end{mapleinput}
\mapleresult
\begin{maplelatex}
\mapleinline{inert}{2d}{`DESOLVII_V5R5 (March-2011)(c) by Dr. K.
T. Vu, Dr. J. Carminati and
Miss. G. Jefferson`;}{%
\maplemultiline{ \mathit{\phantom{xxxxxxxx} DESOLVII\_V5R5\ (March-2011)(c)} \\
\mathit{by\ Dr.\ K.\ T.\ Vu,\ Dr.\ J.\ Carminati\ and\ Miss.\ G.\
Jefferson} }}
\end{maplelatex}
\end{maplegroup}
\begin{maplegroup}
\begin{flushleft}
{\large Definitions of variables and local equation:}
\end{flushleft}
\end{maplegroup}
\begin{maplegroup}
\begin{mapleinput}
\mapleinline{active}{1d}{alias(sigma=(y0,y1,y2,y3,J0,J1,J2)): X:=(y0,y1,y2,y3):
}{}
\end{mapleinput}
\end{maplegroup}
\begin{maplegroup}
\begin{mapleinput}
\mapleinline{active}{1d}{eqn:=diff(J0(X),y0)+diff(J1(X),y1)+diff(J2(X),y2)=0:
}{}
\end{mapleinput}
\end{maplegroup}
\begin{maplegroup}
\begin{flushleft}
{\large Symmetry Algorithm:}\\[0.75em]
\textit{{\large Size of the determining system:}}
\end{flushleft}
\end{maplegroup}
\begin{maplegroup}
\begin{mapleinput}
\mapleinline{active}{1d}{detsys:=gendef([eqn],[J0,J1,J2],[y0,y1,y2,y3]): nops(detsys[1]);
}{}
\end{mapleinput}
\mapleresult
\begin{maplelatex}
\mapleinline{inert}{2d}{24}{\[\displaystyle 24\]}
\end{maplelatex}
\end{maplegroup}
\vspace{-0.75em}
\begin{maplegroup}
\begin{flushleft}
\textit{{\large Solving the determining system:}}
\end{flushleft}
\end{maplegroup}
\begin{maplegroup}
\begin{mapleinput}
\mapleinline{active}{1d}{sym:=pdesolv(op(detsys));
}{}
\end{mapleinput}
\mapleresult
\begin{maplelatex}
\mapleinline{inert}{2d}{}{\[\displaystyle
{\it sym}\, := \,\bigg[\bigg[-{\frac {\partial }{\partial {\it y1}}}{\it F\_6} \left( {\it y0},{\it y1},{\it y2},{\it y3}\\ \mbox{} \right)
-{\frac {\partial }{\partial {\it y2}}}{\it F\_9} \left( {\it y0},{\it y1},{\it y2},{\it y3}\\
\mbox{} \right) -{\it F\_26} \left( {\it y0},{\it y1},{\it y2},{\it y3}\\ \mbox{} \right),
\]}
\end{maplelatex}
\mapleresult
\begin{maplelatex}
\mapleinline{inert}{2d}{}{\[\displaystyle
{\frac {\partial }{\partial {\it y0}}}{\it F\_23} \left( {\it y0},{\it y1},{\it y2},{\it y3}\\
\mbox{} \right)
+{\frac {\partial }{\partial {\it y1}}}{\it F\_24} \left( {\it y0},{\it y1},{\it y2},{\it y3}\\
\mbox{} \right)
+{\frac {\partial }{\partial {\it y2}}}{\it F\_25} \left( {\it y0},{\it y1},{\it y2},{\it y3}\\
\mbox{} \right) \\
\mbox{} \bigg],\,[\,],
\]}
\end{maplelatex}
\mapleresult
\begin{maplelatex}
\mapleinline{inert}{2d}{}{\[\displaystyle
\bigg[\xi_{{{\it y0}}} \left( {\it \sigma}\right)\!=\!{\it F\_3}\\
\mbox{} \left( {\it y0},{\it y1},{\it y2},{\it y3} \right),
\xi_{{{\it y1}}} \left( {\it \sigma}\right)\!=\!{\it F\_6} \left( {\it y0},{\it y1},{\it y2},{\it y3} \right),
\xi_{{{\it y2}}} \left( {\it \sigma}\right)\\
\mbox{}\!=\!{\it F\_9} \left( {\it y0},{\it y1},{\it y2},{\it y3} \right),
\]}
\end{maplelatex}
\mapleresult
\begin{maplelatex}
\mapleinline{inert}{2d}{}{\[\displaystyle
\xi_{{{\it y3}}} \left( {\it \sigma}\right)\!=\!{\it F\_15} \left( {\it y3} \right),
\]}
\end{maplelatex}
\mapleresult
\begin{maplelatex}
\mapleinline{inert}{2d}{}{\[\displaystyle
\eta_{{{\it J0}}} \left( {\it \sigma}\right)\!=\!
{\it F\_26} \left( {\it y0},{\it y1},{\it y2},{\it y3} \right){\it J0}
\!+\!{\it J1}\,{\frac {\partial }{\partial {\it y1}}}{\it F\_3}\\
\mbox{} \left( {\it y0},{\it y1},{\it y2},{\it y3} \right)
\]}
\end{maplelatex}
\mapleresult
\vspace{-0.5em}
\begin{maplelatex}
\mapleinline{inert}{2d}{}{\[\displaystyle
\hspace{1.5cm}+{\it J2}\,{\frac {\partial }{\partial {\it y2}}}{\it F\_3}\\
\mbox{} \left( {\it y0},{\it y1},{\it y2},{\it y3} \right)\!+\!{\it F\_23} \left( {\it y0},{\it y1},{\it y2},{\it y3} \right) \\
\mbox{},
\]}
\end{maplelatex}
\vspace{-0.5em}
\mapleresult
\begin{maplelatex}
\mapleinline{inert}{2d}{}{\[\displaystyle
\eta_{{{\it J1}}} \left( {\it \sigma}\right)\!=\!
{\it J0}\,{\frac {\partial }{\partial {\it y0}}}{\it F\_6} \left( {\it y0},{\it y1},{\it y2},{\it y3} \right) \\
\mbox{}\!+\!{\it F\_26} \left( {\it y0},{\it y1},{\it y2},{\it y3} \right) {\it J1}
\!+\!{\it J2}\,{\frac {\partial }{\partial {\it y2}}}{\it F\_6} \left( {\it y0},{\it y1},{\it y2},{\it y3} \right)
\]}
\end{maplelatex}
\mapleresult
\vspace{-0.25em}
\begin{maplelatex}
\mapleinline{inert}{2d}{}{\[\displaystyle
\hspace{1.5cm}+{\it J1}\,{\frac {\partial }{\partial {\it y1}}}{\it F\_6} \left( {\it y0},{\it y1},{\it y2},{\it y3} \right)
\!-\!{\it J1}\,{\frac {\partial }{\partial {\it y0}}}{\it F\_3}\\
\mbox{} \left( {\it y0},{\it y1},{\it y2},{\it y3} \right)\!+\!{\it F\_24} \left( {\it y0},{\it y1},{\it y2},{\it y3} \right) ,
\]}
\end{maplelatex}
\mapleresult
\begin{maplelatex}
\mapleinline{inert}{2d}{}{\[\displaystyle
\eta_{{{\it J2}}} \left( {\it \sigma}\right) \\
\mbox{}\!=\!
{\it J0}\,{\frac {\partial }{\partial {\it y0}}}{\it F\_9} \left( {\it y0},{\it y1},{\it y2},{\it y3} \right)
\!+\!\,{\it F\_26} \left( {\it y0},{\it y1},{\it y2},{\it y3} \right){\it J2}
\!+\!{\it J2}\,{\frac {\partial }{\partial {\it y2}}}{\it F\_9} \left( {\it y0},{\it y1},{\it y2},{\it y3} \right)
\]}
\end{maplelatex}
\mapleresult
\begin{maplelatex}
\mapleinline{inert}{2d}{}{\[\displaystyle
\hspace{1.5cm}+{\it J1}\,{\frac {\partial }{\partial {\it y1}}}{\it F\_9} \left( {\it y0},{\it y1},{\it y2},{\it y3} \right)
\!-\!{\it J2}\,{\frac {\partial }{\partial {\it y0}}}{\it F\_3}\\
\mbox{} \left( {\it y0},{\it y1},{\it y2},{\it y3} \right)
\!+\!{\it F\_25} \left( {\it y0},{\it y1},{\it y2},{\it y3} \right)\\
\mbox{}\bigg],
\]}
\end{maplelatex}
\mapleresult
\begin{maplelatex}
\mapleinline{inert}{2d}{}{\[\displaystyle
\bigg[{\it F\_3} \left( {\it y0},{\it y1},{\it y2},{\it y3}\\
\mbox{} \right) ,{\it F\_6} \left( {\it y0},{\it y1},{\it y2},{\it y3}\\
\mbox{} \right) ,{\it F\_9} \left( {\it y0},{\it y1},{\it y2},{\it y3}\\
\mbox{} \right) ,{\it F\_23} \left( {\it y0},{\it y1},{\it y2},{\it y3}\\
\mbox{} \right), \\
\mbox{}
\]}
\end{maplelatex}
\mapleresult
\begin{maplelatex}
\mapleinline{inert}{2d}{}{\[\displaystyle
{\it F\_24} \left( {\it y0},{\it y1},{\it y2},{\it y3}\\
\mbox{} \right) ,{\it F\_25} \left( {\it y0},{\it y1},{\it y2},{\it y3}\\
\mbox{} \right) ,{\it F\_26} \left( {\it y0},{\it y1},{\it y2},{\it y3}\\
\mbox{} \right) \\
\mbox{},
{\it F\_15} \left( {\it y3}\\
\mbox{} \right)\bigg]\bigg]
\]}
\end{maplelatex}
\end{maplegroup}
\begin{maplegroup}
\begin{flushleft}
{\large This is exactly the same solution as obtained in Version No.$\,$1.
Just perform the following renaming,}
\end{flushleft}
\end{maplegroup}
%
\begin{maplegroup}
\mapleresult
\begin{maplelatex}
\mapleinline{inert}{2d}{}{\[\displaystyle
{\it F\_3}\rightarrow {\it F\_27}=:{\it \xi 0},\;
{\it F\_6}\rightarrow {\it F\_15}=:{\it \xi 1},\;
{\it F\_9}\rightarrow {\it F\_21}=:{\it \xi 2},\;
{\it F\_15}\rightarrow {\it F\_9}=:{\it \xi 3},\;
\]}
\end{maplelatex}
\mapleresult
\begin{maplelatex}
\mapleinline{inert}{2d}{}{\[\displaystyle
{\it F\_23}\rightarrow {\it F\_44},\;
{\it F\_24}\rightarrow {\it F\_45},\;
{\it F\_25}\rightarrow {\it F\_46},\;
{\it F\_26}\rightarrow {\it F\_47}
\]}
\end{maplelatex}
\end{maplegroup}
\begin{maplegroup}
\begin{flushleft}
{\large then solve for $F_{47}$, and finally identify the three functions $F_{44}$, $F_{45}$ and $F_{46}$ again as}
\end{flushleft}
\end{maplegroup}
\begin{maplegroup}
\mapleresult
\begin{maplelatex}
\mapleinline{inert}{2d}{}{\[\displaystyle
{\it F\_44} \left({\it y0} ,{\it y1} ,{\it y2} ,{\it y3} \right):={\it b0}\\
\mbox{} \left({\it y0} ,{\it y1} ,{\it y2} ,{\it y3} \right)-C \left({\it y3} \right)~{\it J0} \left({\it y0} ,{\it y1} ,{\it y2} ,{\it y3} \right)
\]}
\end{maplelatex}
\mapleresult
\begin{maplelatex}
\mapleinline{inert}{2d}{}{\[\displaystyle
{\it F\_45} \left({\it y0} ,{\it y1} ,{\it y2} ,{\it y3} \right):={\it b1}\\
\mbox{} \left({\it y0} ,{\it y1} ,{\it y2} ,{\it y3} \right)-C \left({\it y3} \right)~{\it J1} \left({\it y0} ,{\it y1} ,{\it y2} ,{\it y3} \right)
\]}
\end{maplelatex}
\mapleresult
\begin{maplelatex}
\mapleinline{inert}{2d}{}{\[\displaystyle
{\it F\_46} \left({\it y0} ,{\it y1} ,{\it y2} ,{\it y3} \right):={\it b2}\\
\mbox{} \left({\it y0} ,{\it y1} ,{\it y2} ,{\it y3} \right)-C \left({\it y3} \right)~{\it J2} \left({\it y0} ,{\it y1} ,{\it y2} ,{\it y3} \right)
\]}
\end{maplelatex}
\end{maplegroup}